\documentclass[11pt,a4paper]{article}

\usepackage[T1]{fontenc}
\usepackage{lmodern}
\usepackage[margin=1in]{geometry}
\usepackage{amsmath,amssymb,amsthm,mathtools}
\usepackage{booktabs}
\usepackage{microtype}
\usepackage{xcolor}
\usepackage{array}
\usepackage{tabularx}
\usepackage{tikz}
\usetikzlibrary{positioning,fit,backgrounds,shapes.geometric}
\usepackage[font=small,labelfont=bf]{caption}
\usepackage[hidelinks]{hyperref}
\hypersetup{pdftitle={What Information Survives Authenticated Byzantine Communication?},pdfauthor={Anonymous Authors}}

\newtheorem{theorem}{Theorem}[section]
\newtheorem{lemma}[theorem]{Lemma}
\newtheorem{proposition}[theorem]{Proposition}
\newtheorem{corollary}[theorem]{Corollary}
\theoremstyle{definition}
\newtheorem{definition}[theorem]{Definition}
\newtheorem{example}[theorem]{Example}
\newtheorem{scope}[theorem]{Scope}
\theoremstyle{remark}
\newtheorem{remark}[theorem]{Remark}

\newcommand{\cA}{\mathcal{A}}
\newcommand{\cC}{\mathcal{C}}
\newcommand{\cE}{\mathcal{E}}
\newcommand{\bP}{\mathbb{P}}
\newcommand{\bE}{\mathbb{E}}
\newcommand{\Comp}{\operatorname{Comp}}
\newcommand{\View}{\operatorname{View}}
\newcommand{\sid}{\mathsf{sid}}
\newcommand{\botmsg}{\bot}
\newcommand{\dist}{\operatorname{dist}}
\newcommand{\Wone}{W[1]}
\newcommand{\LLR}{\operatorname{LLR}}
\newcommand{\orcidicon}[1]{\href{https://orcid.org/#1}{\includegraphics[height=1.7ex]{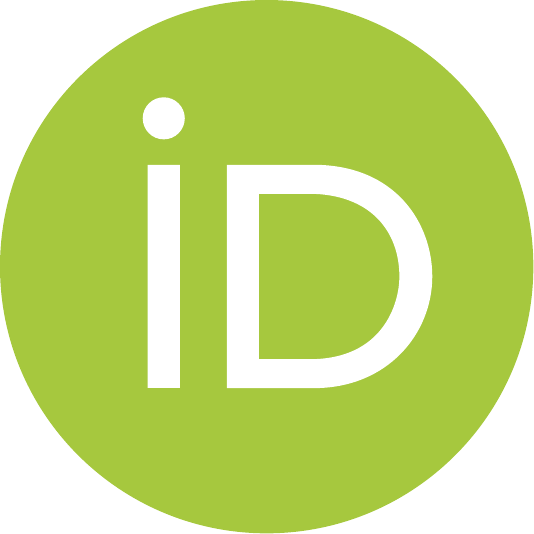}\,#1}}
\title{What Information Survives Authenticated Byzantine Communication?}
\author{
  Arnab Mallick\\
  \small Centre for Development of Advanced Computing, Hyderabad, India\\
  \small \texttt{arnabm@cdac.in} \\
  \small \orcidicon{0009-0008-5458-6429}
  \and
  Indraveni Chebolu\\
  \small Centre for Development of Advanced Computing, Hyderabad, India\\
  \small \texttt{indravenik@cdac.in}\\
  \small \orcidicon{0009-0009-6128-6253}
}
\date{Full version}

\begin{document}

\maketitle

\begin{abstract}
A correct process in a Byzantine network may receive many authenticated
statements and still be unable to tell which statistical facts are true:
a signature authenticates the speaker, not the speaker's honesty.  We study
this distinction for distributed binary hypothesis testing.  Each process
obtains private i.i.d. samples from a process-specific distribution, up to
$f$ non-target processes are statically Byzantine, communication is synchronous
and authenticated, and a fixed correct target must infer the hidden hypothesis.
Our goal is the best worst-case exponential decay rate of the target's error
probability over all finite-round interactive protocols.

For two candidate fault sets $F_0,F_1$, form an \emph{authenticated pair graph}
by deleting the processes faulty in both candidate worlds and every edge whose
endpoints are faulty exclusively in opposite worlds.  Let $S_v(F_0,F_1)$ be
the common-correct sources in the target's remaining component.  We prove the
exact characterization
\[
  \mathcal E_v^*
  =
  \min_{|F_0|,|F_1|\le f}\mathcal C_{S_v(F_0,F_1)},
\]
where $\mathcal C_S$ is the Chernoff information of the product experiment
carried by sources $S$.  The converse constructs one two-world channel from
that residual experiment to the target view.  Achievability uses a standard
authenticated full-information flood, together with a cross-world
causal-path certification lemma: different edges of a pair-graph path may be
certified in different candidate worlds, yet one immutable signed target
transcript exposes a genuine causal chain.  Thus the same pair object exactly
governs impossibility and achievability.

The characterization is information-theoretic: message payloads and local
computation are unrestricted.  Succinct bit complexity and efficient decoding
are separate algorithmic questions.  On pair-connected graphs it reduces to
deletion of any $2f$
non-target sources, recovering the familiar centralized $2f$-type phenomenon
as a special case.  On incomplete graphs the surviving experiment depends
jointly on topology, the target, the two fault explanations, and source
informativeness.  As one application, interval observations of a latent event
order yield an exact temporal boundary-cut formula and, under a nondegenerate
homogeneous trace channel, an exact Byzantine error exponent.
\end{abstract}

\section{Introduction}
\label{sec:intro}

\paragraph{The problem.}
There is an unknown bit $\theta\in\{0,1\}$.  Process $i$ receives $T$ noisy
private samples drawn from a distribution $P_{i,\theta}$, and the processes may
communicate over a known graph $G=(V,E)$.  Up to $f$ non-target processes are
Byzantine: they may omit messages, send different statements to different
neighbors, coordinate globally, and lie using their own identities.  Messages
are authenticated, so a Byzantine process cannot impersonate a correct one,
but authentication does not make a Byzantine statement truthful.  A fixed
correct target $v$ must output an estimate of $\theta$.

For a protocol $\Pi_T$, let $R_{v,T}(\Pi_T)$ be the target's worst-case error
probability, maximized over the hypothesis, every admissible static fault set,
and every Byzantine strategy.  The central quantity of this paper is
\[
  \mathcal E_v^*
  =
  \sup_{(\Pi_T)}\liminf_{T\to\infty}
  -\frac1T\log R_{v,T}(\Pi_T),
\]
the best exponential rate at which that worst-case error can vanish.  This is
\emph{target-specific inference}.  It is not Byzantine agreement and it is not
the reliable delivery of a particular source's message.

\paragraph{Why this is a distributed-computing question.}
Classical Byzantine detection and robust-inference work already studies
compromised sensors, fake reporting distributions, alternative plausible
traitor sets, Chernoff error exponents, and $2f$-type losses
\cite{KosutTong2008Source,KosutTong2008CEO,MaranoMattaTong2009,RenYanMo2018}.
These statistical ingredients are established.  Authenticated communication
work, in turn, characterizes when specific payloads can be transported or
agreed upon using paths, cuts, signatures, and trust structure
\cite{DolevStrong1983,BeimelFranklin1999,Pagourtzis2017,ChotkanEtAl2024}.
Our contribution lies at the intersection of these two literatures:

\begin{quote}
\emph{Given arbitrary authenticated interaction on an incomplete graph, which
statistical experiment can a Byzantine adversary force a correct target to
retain, and what is the resulting optimal testing exponent?}
\end{quote}

The distinction matters because statistical evidence can substitute across
sources.  A particular source may be impossible to deliver reliably while
other sources carry equivalent evidence, so reliable communication is stronger
than inference requires.  Conversely, two graphs with the same classical
connectivity can induce different inference exponents when the informative
sources occupy different vertices.  The relevant object must therefore couple
network topology with source identity and statistical informativeness.

\paragraph{Two fault explanations.}
An impossibility argument must compare two plausible executions: under
hypothesis $0$, fault set $F_0$ may explain the target's view, while under
hypothesis $1$, another set $F_1$ may explain the same view.  Write
$B=F_0\cap F_1$, $A_0=F_0\setminus F_1$, and $A_1=F_1\setminus F_0$.
A process in $B$ is Byzantine in both worlds and can suppress its behavior in
both.  More subtly, an edge between $A_0$ and $A_1$ can be controlled in either
world, because one of its endpoints is Byzantine in each execution.  This
leads to the authenticated pair graph
\begin{equation}
\label{eq:intro-pairgraph}
H(F_0,F_1)=G-(F_0\cap F_1)-E(F_0\setminus F_1,F_1\setminus F_0).
\end{equation}
Let $K_v(F_0,F_1)$ be the target's connected component in this graph and
\[
  S_v(F_0,F_1)=K_v(F_0,F_1)\setminus(F_0\cup F_1).
\]
These are exactly the common-correct sources that the two candidate Byzantine
explanations cannot remove from the target's residual experiment.

For a source set $S$, let $\mathcal C_S$ denote the centralized Chernoff
information of the product observations of $S$.  Operationally,
$\mathcal C_S$ is the optimal exponential rate at which a centralized observer
that is handed exactly those samples can distinguish the two hypotheses.  Our
main theorem is
\begin{equation}
\label{eq:intro-main}
\mathcal E_v^*
=
\Lambda_v
:=
\min_{\substack{F_0,F_1\subseteq V\setminus\{v\}\\|F_0|,|F_1|\le f}}
\mathcal C_{S_v(F_0,F_1)}.
\end{equation}
Thus the distributed part of the theorem identifies \emph{which experiment
must survive}; the statistical part evaluates how informative that experiment
is.

\paragraph{Why exactness is nontrivial.}
For the converse, we show that for every $F_0,F_1$ the adversary can realize
the two hypotheses through one common stochastic channel whose only
hypothesis-dependent input is the observation block of $S_v(F_0,F_1)$.  No
interactive protocol can create statistical information that is absent from
that source experiment.

The reverse direction is more delicate.  We run a standard authenticated
full-information causal flood.  A path in $H(F_0,F_1)$ need not consist of
correct processes in either candidate world: one edge may be certifiable only
in world $0$, and the next only in world $1$.  Lemma~\ref{lem:causal-path-exposure}
proves by backward induction that the certifying world can switch only at a
vertex correct in both worlds.  Fresh-session authentication and the
one-state-per-round rule then force each recovered signed state to be the same
concrete object in every candidate execution in which its signer is correct.
Consequently, one target transcript contains the authentic observations of
every source in $S_v(F_0,F_1)$, without path-indexed communication.

\paragraph{Scope of the characterization.}
On an $f$-pair-connected graph, including the complete graph, the formula
collapses to
\[
  \Lambda_v
  =
  \min_{\substack{U\subseteq V\setminus\{v\}\\|U|\le 2f}}
  \mathcal C_{V\setminus U}.
\]
Hence the familiar $2f$-source loss is recovered as a special case.  On incomplete graphs, the pair
component can remove additional sources, and which sources disappear matters
statistically.  The characterization imposes no separate numerical resilience
condition on $f$ beyond the requirement that the target be correct.  Feasibility
is instead determined by the residual statistical information in every
admissible pair graph.  In particular, an informative target always retains its
own evidence.

The theorem is information-theoretic.  The achieving protocol uses standard
full-information flooding and schedules at most $2(n-1)|E|$ correct-process
transmissions, while signed causal states may be large and compatibility
testing may be computationally expensive.  The achievability result is a
sufficiency statement: a single causal transcript contains exactly the evidence
needed to attain the all-protocol converse.  Succinct bit complexity and
efficient decoding remain separate algorithmic questions.

\paragraph{Further results.}
Section~\ref{sec:temporal} gives one concrete application.  Reporters observe
contiguous windows of a latent event order.  We show that the minimum number of
reporters whose local traces must change to realize a prescribed Kendall
distortion has an exact boundary-cut form.  Under pair-connectivity and a
nondegenerate homogeneous trace-change channel this becomes the closed-form
Byzantine exponent $c[d_{\rm rel}^{\pi}(\Gamma)-2f]_+$.  The appendices show
that $2f$-vertex-fault-tolerant spanners preserve the pair information, prove
W[1]-hardness of evaluating $\Lambda_v$ exactly even for a restricted source
family, record the model assumptions and scope, and distinguish finite-hypothesis
identification from baseline-versus-composite testing.

The rest of the paper proceeds as follows.  Section~\ref{sec:model} formalizes
the model and testing objective; Section~\ref{sec:pair-graph} defines the pair
graph and states the exact theorem; Section~\ref{sec:proof-main} gives the core
converse and achievability arguments; Section~\ref{sec:temporal} develops the
order application; and Section~\ref{sec:related} positions the result against
Byzantine inference and authenticated communication.  Full proofs and
secondary consequences follow in the appendices.

\section{Model and Problem Formulation}
\label{sec:model}

We study distributed binary hypothesis testing in a synchronous network with
static Byzantine faults and transferable authentication.  The formulation is
information-theoretic: signatures are ideal symbolic objects, message payloads
and local computation are unrestricted, and the performance criterion is the
best asymptotic testing exponent attainable by finite-round interactive
protocol families.  For each sample size $T$, a protocol may use any finite
number $R_T$ of rounds, possibly depending on $T$; the converse covers every
such finite $R_T$, while the achieving protocol uses the uniform bound
$R_T=n-1$.  We record the number of scheduled transmissions for the achieving
flood, counting one send over one oriented edge in one designated round as one
transmission regardless of payload length.

\subsection{Network and faults}

Let
\[
    G=(V,E), \qquad |V|=n,
\]
be a known undirected communication graph.  We fix a distinguished target
process $v\in V$ whose decision quality we study.  A \emph{fault set} is a set
\[
    F\subseteq V\setminus\{v\}, \qquad |F|\le f.
\]
Processes in $F$ are Byzantine and all other processes are correct.  Faults are
\emph{static}: $F$ is chosen before the random observation blocks are drawn and
remains fixed throughout the execution.  The Byzantine strategy may subsequently
learn the realized correct observations.  The target $v$ is correct by
definition.

Communication proceeds in synchronous rounds.  In each round, every correct
process first computes its outgoing messages from its private input, private
randomness, and all messages delivered in previous rounds.  The adversary may
then observe the correct round-$r$ outgoing messages before choosing the
round-$r$ messages sent by Byzantine processes.  Thus the model permits a
\emph{rushing} Byzantine adversary.  Every message sent over an edge between two
correct processes in round $r$ is delivered by the end of round $r$.
Byzantine processes may omit messages, send malformed messages, send different
messages to different neighbors, and otherwise deviate arbitrarily.

\subsection{Transferable authentication}

We use an ideal transferable-signature abstraction.  A signed object is written
$[m]_i$ and has the following properties.

\begin{enumerate}
\renewcommand{\labelenumi}{(S\arabic{enumi})}
    \item Only process $i$ can create a new object carrying a valid signature
          of $i$.
    \item Any process can forward an existing signed object unchanged, and any
          recipient can verify the signer and payload.
    \item A Byzantine process $i$ may sign arbitrary payloads under identity
          $i$.
    \item Signatures bind the signer, session identifier, round/stage tags, and
          payload.  The notation $[m]_i$ denotes the concrete valid signed
          object produced in that execution; deterministic signature bytes are
          not required.
    \item The session identifier is fresh.  Before round~$0$, the adversary
          holds no valid same-session object signed by a process that is correct
          in the candidate execution; objects from other sessions never verify
          under $\sid$.  Thus every valid same-session certificate of a correct
          signer originates in that signer's current execution.
\end{enumerate}

The bracket notation is canonical only at the level of the authenticated
statement, not at the level of signature-generation bits.  If an ideal signing
procedure uses private coins, include those coins in the signer's
hypothesis-independent random tape.  Conditioned on that tape, the concrete
signed object is determined; in the converse, these signing coins are simply
part of the common auxiliary randomness $Z$.  In achievability, we use only the
following fact: the exact signed object actually sent on a correct-to-correct
edge can be forwarded and verified, and cannot have been forged under a
correct signer's identity.  We never require a second signing invocation on the
same payload to reproduce the same bit string.

Thus hypothesis-independent signature-generation randomness is absorbed into
the same post-processing kernel and cannot change the statistical exponent.  A computational
instantiation can be treated separately by choosing a security parameter whose
forgery probability is negligible on the statistical error scale of interest.

\subsection{Private observations}

The global hypothesis is
\[
    \theta\in\{0,1\}.
\]
For each process $i\in V$, let $\mathcal{X}_i$ be a finite observation alphabet
and let $P_{i,0}$ and $P_{i,1}$ be probability distributions on
$\mathcal{X}_i$.  Process $i$ receives a block
\[
    X_i^T=(X_{i,1},\ldots,X_{i,T}),
\]
where, conditioned on $\theta$,
\[
    X_{i,1},\ldots,X_{i,T}\stackrel{\mathrm{iid}}{\sim}P_{i,\theta}.
\]
Observation blocks of distinct processes are conditionally independent given
$\theta$:
\begin{equation}
    \label{eq:product-observation-law}
    P_{\theta}^{(T)}
    =
    \bigotimes_{i\in V} P_{i,\theta}^{\otimes T}.
\end{equation}
This conditional-independence assumption is used essentially in the converse;
see Remark~\ref{rem:correlation}.

For notational convenience in the main proof, we assume that $P_{i,0}$ and
$P_{i,1}$ have common support.  This avoids infinite log-likelihood ratios.
The statement extends to unequal supports with the usual conventions and only
becomes easier when an observation rules out one hypothesis exactly.

\subsection{Protocols and adversaries}

For each sample size $T$, a protocol $\Pi_T$ terminates after a deterministic
finite number $R_T<\infty$ of synchronous rounds.  Correct process behavior may
be randomized.  The private random tapes of correct processes are mutually
independent and independent of all observation blocks.

The Byzantine adversary may know the true hypothesis $\theta$, the entire
collection of correct observations, the protocol description, and the complete
execution history.  It may coordinate all faulty processes, use an arbitrary
private random tape, and rush within rounds as described above.  The adversarial
random tape is independent of the source draws before the execution but may be
used adaptively after observations and messages are revealed.  We write $\cA$
for an admissible Byzantine strategy.  Allowing this randomization only
strengthens the adversary; Appendix~\ref{subsec:converse} also explains why the
same converse follows under a deterministic-adversary convention by fixing a
seed with at least the averaged error.

The target's complete local information at termination---its observation block,
private randomness, and received authenticated messages---is denoted by
\[
    \View_v^{\theta,F,\cA}(\Pi_T).
\]
The protocol outputs an estimate
\[
    \widehat\theta_v\in\{0,1\}.
\]

\subsection{Minimax risk and error exponent}

\begin{definition}[Target minimax risk]
\label{def:risk}
For a protocol $\Pi_T$, define
\begin{equation}
    \label{eq:minimax-risk}
    R_{v,T}(\Pi_T)
    =
    \max_{\theta\in\{0,1\}}
    \max_{\substack{F\subseteq V\setminus\{v\}\\ |F|\le f}}
    \sup_{\cA}
    \bP_{\theta,F,\cA}
    \bigl[\widehat\theta_v\neq\theta\bigr].
\end{equation}
\end{definition}

Thus $R_{v,T}(\Pi_T)$ is the target error probability under the worst choice of
true hypothesis, admissible static fault set, and Byzantine strategy.

For a protocol family $\boldsymbol\Pi=(\Pi_T)_{T\ge1}$, define its target error
exponent by
\begin{equation}
    \label{eq:protocol-exponent}
    \cE_v(\boldsymbol\Pi)
    =
    \liminf_{T\to\infty}
    -\frac{1}{T}\log R_{v,T}(\Pi_T).
\end{equation}
The optimal exponent is
\begin{equation}
    \label{eq:optimal-exponent}
    \cE_v^\star
    =
    \sup_{\boldsymbol\Pi}\cE_v(\boldsymbol\Pi).
\end{equation}
Throughout the asymptotic limit, the network $G$, $n$, and $f$ are fixed while
$T\to\infty$.

\subsection{Chernoff information of a source set}

For a distributed-computing reader, $\mathcal C_S$ can be read operationally as
follows: if a centralized observer were handed exactly the observations of the
sources in $S$, then $\mathcal C_S$ is the best exponential rate at which its
binary testing error can decay with the number of samples.  Larger values mean
that the surviving sources distinguish the hypotheses more strongly;
$\mathcal C_S=0$ means there is no exponential-scale separation from those
sources alone.  Formally, for $S\subseteq V$, define
\begin{align}
    \label{eq:chernoff-set}
    \cC_S
    &:=
    \max_{s\in[0,1]}
    \left
    \{
        -\sum_{i\in S}
        \log
        \left(
            \sum_{x\in\mathcal{X}_i}
            p_{i,0}(x)^{1-s}p_{i,1}(x)^s
        \right)
    \right\}.
\end{align}
This is the Chernoff information between the product measures
$\bigotimes_{i\in S}P_{i,0}$ and $\bigotimes_{i\in S}P_{i,1}$.  The common tilt
parameter $s$ is important when the sources are heterogeneous.

By the classical Chernoff theorem~\cite{Chernoff1952}, if a centralized tester receives all $T$
i.i.d. samples from the processes in $S$, then its optimal equal-prior Bayes
error $P_{e,S}^\star(T)$ satisfies
\begin{equation}
    \label{eq:chernoff-theorem}
    \lim_{T\to\infty}
    -\frac{1}{T}\log P_{e,S}^\star(T)
    =
    \cC_S.
\end{equation}

\begin{remark}[Why independence is explicit]
\label{rem:correlation}
The converse below constructs one common stochastic channel from the observations
of a set of common-correct processes to the target view.  The construction uses
independent auxiliary observations for processes that are correct in only one of
two candidate fault worlds.  With arbitrary cross-process correlations, those
auxiliary variables may have hypothesis-dependent correlations with the
common-correct observations, and the same channel need not exist.  The theorem
therefore assumes conditional independence; arbitrary cross-process
correlations require a different coupling or joint information quantity.
\end{remark}

\section{Authenticated Pair Graphs and the Exact Information Parameter}
\label{sec:pair-graph}

The key object is induced by \emph{two} candidate Byzantine explanations.  A
process faulty under both explanations can suppress information in both worlds.
If two adjacent processes are faulty exclusively in opposite worlds, the two
worlds can hand a simulation across their edge because one endpoint is Byzantine
in either execution.  Authentication prevents such a handoff across a process
that is correct in both worlds.

\subsection{The authenticated pair graph}

Fix two candidate fault sets
\[
    F_0,F_1\subseteq V\setminus\{v\},
    \qquad
    |F_0|,|F_1|\le f.
\]
Write
\begin{equation}
    \label{eq:A0A1B}
    B:=F_0\cap F_1,
    \qquad
    A_0:=F_0\setminus F_1,
    \qquad
    A_1:=F_1\setminus F_0.
\end{equation}

\begin{definition}[Authenticated pair graph]
\label{def:pair-graph}
The authenticated pair graph associated with $(F_0,F_1)$ is
\begin{equation}
    \label{eq:pair-graph}
    H(F_0,F_1)
    :=
    G-B-E(A_0,A_1),
\end{equation}
where $G-B$ deletes every vertex in $B$ and its incident edges, and
$E(A_0,A_1)$ denotes all original graph edges with one endpoint in $A_0$ and
one endpoint in $A_1$.
\end{definition}

Let
\begin{equation}
    \label{eq:Kv}
    K_v(F_0,F_1)
    :=
    \Comp_v\bigl(H(F_0,F_1)\bigr)
\end{equation}
be the connected component containing the target $v$.  Define
\begin{equation}
    \label{eq:Sv}
    S_v(F_0,F_1)
    :=
    K_v(F_0,F_1)\setminus(F_0\cup F_1).
\end{equation}
Thus $S_v(F_0,F_1)$ consists of the processes that are correct in both candidate
worlds and remain connected to $v$ after exactly the two-world simulation
handoffs are removed.  In particular, $v\in S_v(F_0,F_1)$.

The pair graph is symmetric in its two arguments:
\[
    H(F_0,F_1)=H(F_1,F_0),
    \qquad
    S_v(F_0,F_1)=S_v(F_1,F_0).
\]

\begin{figure}[t]
\centering
\begin{tikzpicture}[x=1.05cm,y=1.00cm,>=latex,
  cc/.style={circle,draw,minimum size=5.6mm,inner sep=0pt},
  a0/.style={regular polygon,regular polygon sides=3,draw,fill=gray!20,minimum size=7mm,inner sep=0pt},
  a1/.style={rectangle,draw,fill=gray!20,minimum size=5.6mm,inner sep=0pt},
  both/.style={circle,draw,fill=black,text=white,minimum size=5.8mm,inner sep=0pt},
  target/.style={circle,draw,double,minimum size=6.2mm,inner sep=0pt},
  removed/.style={densely dashed,gray,thick}]
  \node[cc] (s) at (-2.5,0.7) {$s$};
  \node[target] (v) at (-0.8,0) {$v$};
  \node[cc] (c) at (1.0,0.6) {$c$};
  \node[a0] (x) at (2.8,0.9) {$x$};
  \node[a1] (y) at (4.7,0.1) {$y$};
  \node[cc] (t) at (6.4,0.7) {$t$};
  \node[both] (b) at (0.9,-1.5) {$b$};
  \node[cc] (z) at (2.8,-1.8) {$z$};
  \draw (s)--(v)--(c)--(x);
  \draw[removed] (x)--node[above,sloped,font=\scriptsize]{deleted $A_0$--$A_1$ edge}(y);
  \draw (y)--(t);
  \draw[removed] (v)--(b);
  \draw[removed] (b)--(z);
  \begin{scope}[on background layer]
    \node[draw=gray!55,rounded corners,inner sep=5pt,fit=(s)(v)(c)(x),label={[font=\scriptsize]above:$K_v(F_0,F_1)$}] {};
  \end{scope}
  \node[font=\scriptsize,anchor=west] at (-2.8,-2.45) {$\circ$ common-correct \quad $\triangle$ $A_0=F_0\setminus F_1$ \quad $\square$ $A_1=F_1\setminus F_0$ \quad $\bullet$ $B=F_0\cap F_1$};
\end{tikzpicture}
\caption{The authenticated pair graph removes common faults and edges joining
opposite-exclusive faults.  Here $t$ is common-correct but lies outside the
target component because the $A_0$--$A_1$ handoff edge is deleted; $z$ is cut
off by the common fault $b$.  The statistical experiment forced to survive is
carried only by the common-correct vertices inside the shaded component, namely
$S_v(F_0,F_1)$.}
\label{fig:pair-graph}
\end{figure}
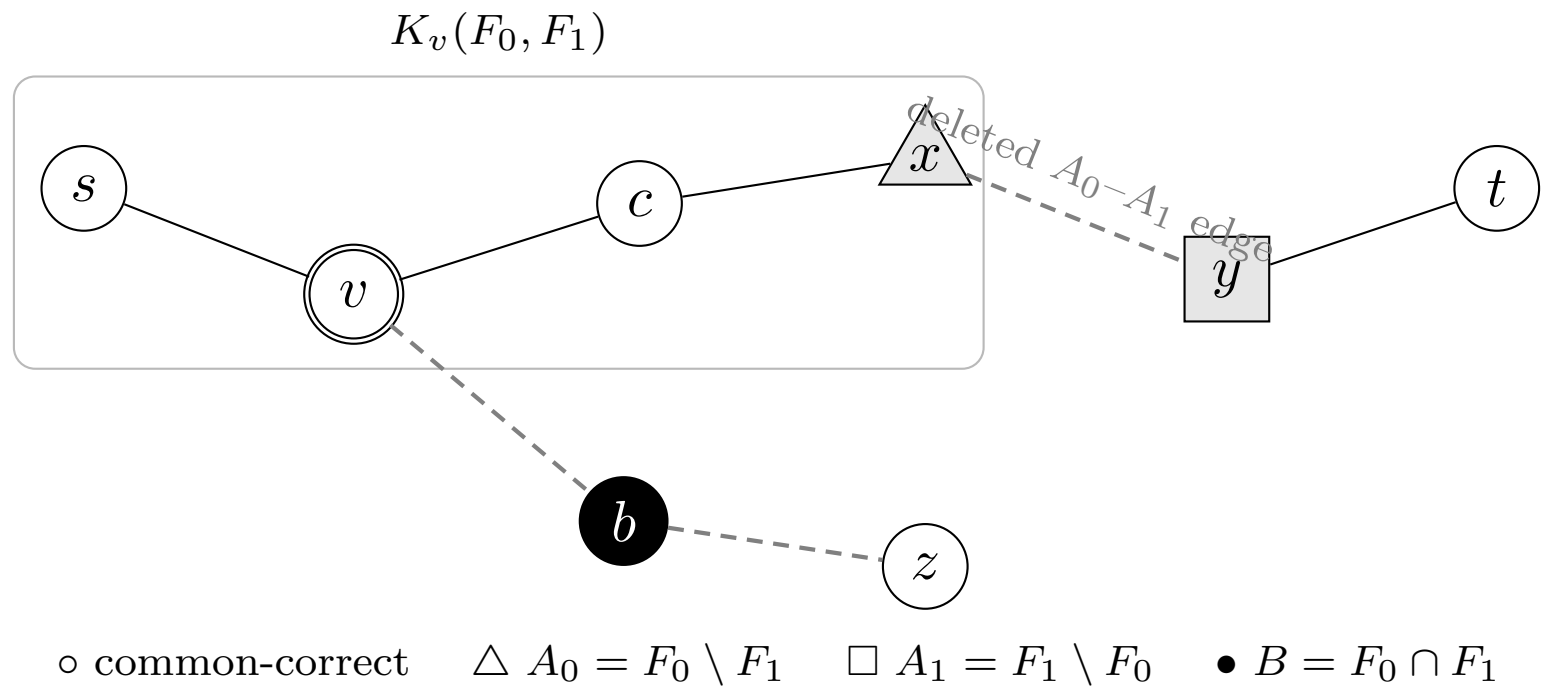

\subsection{The pair-information parameter}

\begin{definition}[Authenticated pair information]
\label{def:Lambda}
Define
\begin{equation}
    \label{eq:Lambda}
    \Lambda_v
    :=
    \min_{\substack{
        F_0,F_1\subseteq V\setminus\{v\}\\
        |F_0|,|F_1|\le f
    }}
    \cC_{S_v(F_0,F_1)}.
\end{equation}
\end{definition}

The minimization has a direct two-world interpretation.  Under hypothesis $0$,
processes in $F_0$ may be Byzantine; under hypothesis $1$, processes in $F_1$
may be Byzantine.  The processes in $S_v(F_0,F_1)$ are precisely the
common-correct sources whose hypothesis-dependent observations cannot be
removed from the target's two-world experiment by the graph/simulation
structure.  The quantity $\Lambda_v$ chooses the most confusable pair of fault
explanations and measures the Chernoff information that survives that pair.

\begin{theorem}[Exact authenticated Byzantine inference]
\label{thm:main}
Under the model of Section~\ref{sec:model},
\begin{equation}
    \label{eq:main-theorem}
    \cE_v^\star
    =
    \Lambda_v
    =
    \min_{\substack{
        F_0,F_1\subseteq V\setminus\{v\}\\
        |F_0|,|F_1|\le f
    }}
    \cC_{S_v(F_0,F_1)}.
\end{equation}
Moreover, the standard causal-state flood described in
Section~\ref{sec:proof-main} attains this exponent for every sample size $T$.
It terminates after $R=n-1$ synchronous rounds, schedules one state transmission
per oriented edge per round, and therefore sends at most
\begin{equation}
    \label{eq:main-message-bound}
    2(n-1)|E|
\end{equation}
correct-process transmissions.  Its minimax risk satisfies
\begin{equation}
    \label{eq:main-risk-bound}
    R_{v,T}
    \le
    \min\{1,N_f e^{-T\Lambda_v}\},
    \qquad
    N_f:=\sum_{j=0}^{f}\binom{n-1}{j}.
\end{equation}
\end{theorem}

The equality is over the entire class of finite-round interactive protocols;
the flood is not tailored to the minimizing pair $(F_0,F_1)$.  The
bound~\eqref{eq:main-message-bound} counts scheduled transmissions in the
unbounded-payload model of Section~\ref{sec:model}.  The protocol itself is
standard full-information flooding; the new statement is that one such
transcript is sufficient for the exact pair-information bound, without
path-indexed protocol state or path-specific communication.

\begin{definition}[$f$-pair-connectivity at a target]
\label{def:pair-connected-target}
The graph $G$ is \emph{$f$-pair-connected at $v$} if for every
$F_0,F_1\subseteq V\setminus\{v\}$ with $|F_0|,|F_1|\le f$,
\begin{equation}
    \label{eq:pair-connected-target}
    K_v(F_0,F_1)=V\setminus(F_0\cap F_1).
\end{equation}
Equivalently, every process not deleted as a common fault remains in the
target component of every admissible authenticated pair graph.
\end{definition}

\begin{corollary}[Pair-connected networks reduce to $2f$ source erasures]
\label{cor:pair-connected-general}
If $G$ is $f$-pair-connected at $v$, then
\begin{equation}
    \label{eq:pair-connected-lambda}
    \boxed{
    \Lambda_v
    =
    \min_{\substack{U\subseteq V\setminus\{v\}\\|U|\le 2f}}
    \cC_{V\setminus U}.
    }
\end{equation}
In particular, a complete communication graph is $f$-pair-connected at every
correct target for every feasible $f$.
\end{corollary}

\begin{proof}
Pair-connectivity gives $S_v(F_0,F_1)=V\setminus(F_0\cup F_1)$, and the possible
unions $F_0\cup F_1$ are exactly the subsets of $V\setminus\{v\}$ of size at
most $2f$.  In a complete graph, $v$ remains adjacent to every non-common-fault
vertex.
\end{proof}

\begin{corollary}[The target's own information is protected]
\label{cor:target-protected}
For every graph and fault budget,
\[
    \Lambda_v\ge \mathcal C_{\{v\}}.
\]
Hence an informative target has a strictly positive exponent regardless of
network topology or Byzantine behavior.
\end{corollary}
\begin{proof}
The target is excluded from every admissible fault set and belongs to every
$S_v(F_0,F_1)$.  Chernoff information is monotone under adding independent
coordinates, so each term minimized in~\eqref{eq:Lambda} is at least
$\mathcal C_{\{v\}}$.
\end{proof}

\begin{figure}[t]
\centering
\begin{tikzpicture}[
  x=0.68cm,y=0.68cm,
  every node/.style={circle,draw,minimum size=5.2mm,inner sep=1.0pt,font=\scriptsize},
  target/.style={circle,draw,double,minimum size=5.8mm,inner sep=1.0pt,font=\scriptsize\bfseries},
  lab/.style={draw=none,rectangle,font=\scriptsize,inner sep=0pt}
]
  \node[target] (vA) at (0,0) {$v_1$};
  \node (aA) at (1.2,0.8) {$a_1$};
  \node (bA) at (2.4,0) {$b_8$};
  \node (cA) at (2.5,1.6) {$c_8$};
  \node (dA) at (3.7,1.6) {$d_8$};
  \node (eA) at (4.9,1.6) {$e_1$};
  \draw (vA)--(aA)--(bA)--(vA); \draw (aA)--(cA)--(dA)--(eA);
  \node[lab] at (2.45,-0.55) {A: $\Lambda_v=9$};
  \begin{scope}[xshift=6.4cm]
  \node[target] (vB) at (0,0) {$v_1$};
  \node (aB) at (1.2,0.8) {$a_1$};
  \node (bB) at (2.4,0) {$b_1$};
  \node (cB) at (2.5,1.6) {$c_8$};
  \node (dB) at (3.7,1.6) {$d_8$};
  \node (eB) at (4.9,1.6) {$e_8$};
  \draw (vB)--(aB)--(bB)--(vB); \draw (aB)--(cB)--(dB)--(eB);
  \node[lab] at (2.45,-0.55) {B: $\Lambda_v=2$};
  \end{scope}
\end{tikzpicture}
\caption{Same communication graph and $f=1$, different pair information.
Subscripts are one-source Chernoff weights.  The pair
$F_0=F_1=\{a\}$ leaves $\{v,b\}$ in both instances, but moving the high-value
sources changes the exact exponent from $9$ to $2$.}
\label{fig:weighted-pair-example}
\end{figure}
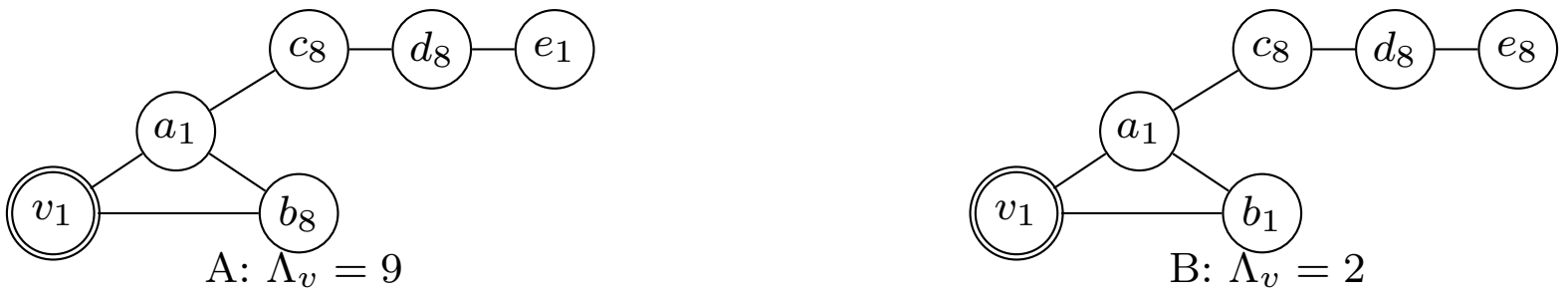

\begin{example}[Same topology and resilience, different surviving information]
\label{ex:weighted-same-topology}
Let $f=1$ and use the same six-process graph in both panels of
Figure~\ref{fig:weighted-pair-example}: the triangle $v-a-b-v$ with tail
$a-c-d-e$.  Thus every unweighted connectivity, cut, path, and resilience
property is identical.  Choose symmetric binary experiments
$P_{i,0}=\operatorname{Bern}(p_i)$,
$P_{i,1}=\operatorname{Bern}(1-p_i)$ whose common Chernoff-optimal tilt is
$s=1/2$, with displayed one-source weights $w_i\in\{1,8\}$.  Hence
$\mathcal C_S=\sum_{i\in S}w_i$.

\emph{Instance A.}
The weights are $(1,1,8,8,8,1)$ on $(v,a,b,c,d,e)$.
The common fault $F_0=F_1=\{a\}$ leaves $S_v=\{v,b\}$, giving
$\Lambda_v\le9$.  For the reverse bound, if $b\notin F_0\cup F_1$ then $b$
survives on edge $v-b$; if $b$ is excluded but $c$ is not, then
$a\notin F_0\cap F_1$ because $f=1$, and the retained path $v-a-c$ exposes
$c$; and if both $b,c$ are the two opposite-exclusive faults, $d$ survives on
$v-a-c-d$.  Thus $\Lambda_v=9$.

\emph{Instance B.}
The same weight multiset is moved so that $w_v=w_a=w_b=1$ and $w_c=w_d=w_e=8$.
The same common fault at $a$ gives $\Lambda_v\le2$.  For the reverse bound,
if $a$ is a common fault then $b$ remains common-correct on edge $v-b$; if $b$
is a common fault then $a$ remains common-correct on edge $v-a$; if $a,b$ are
the two opposite-exclusive faults then $c$ remains common-correct in the target
component through $v-a-c$; and every other admissible pair leaves at least one
of $a,b$ common-correct and adjacent to $v$.  Thus $\Lambda_v\ge2$.  Therefore
\[
    \Lambda_v^{\mathrm A}=9\qquad\text{and}\qquad\Lambda_v^{\mathrm B}=2.
\]
Instance A also lies strictly between the two immediate benchmarks from the
main theorem.  Its target has weight $\mathcal C_{\{v\}}=1$, while deleting the
two heaviest non-target sources from total weight $27$ leaves weight $11$.
Hence
\begin{equation}
\label{eq:example-intermediate}
    1=\mathcal C_{\{v\}}
    <\Lambda_v^{\mathrm A}=9
    <\min_{\substack{U\subseteq V\setminus\{v\}\\|U|\le2}}
      \mathcal C_{V\setminus U}=11.
\end{equation}
The topology is unchanged between A and B; only the identities of
informative sources move.
\end{example}

\begin{remark}[Target-specific inference versus agreement]
\label{rem:target-vs-agreement}
We impose no agreement requirement across correct targets, and $\Lambda_v$ may
vary with $v$.  Requiring all correct processes to output one common hypothesis
is a different problem combining statistical distinguishability with
authenticated Byzantine agreement.  The targetwise $\Lambda_v$ values are
necessary information constraints but do not characterize that optimum; we
leave the agreement-constrained problem open.
\end{remark}

\section{Proof Overview of the Exact Pair-Graph Characterization}
\label{sec:proof-main}

The equality in Theorem~\ref{thm:main} has two logically independent parts.
The converse identifies a statistical experiment that every protocol can be
forced to factor through.  Achievability shows that one authenticated causal
history exposes exactly the common-correct sources that survive every such
factorization.  Complete proofs, including randomized protocols and rushing,
appear in Appendix~\ref{app:full-proof}.

\subsection{Converse: one common channel for two Byzantine worlds}
Fix $F_0,F_1$ and abbreviate
$B=F_0\cap F_1$, $A_0=F_0\setminus F_1$, $A_1=F_1\setminus F_0$,
$K=K_v(F_0,F_1)$, and $S=S_v(F_0,F_1)$.  Every original edge leaving $K$ is
absent from the pair graph.  Consequently it either meets a common fault in
$B$ or is an $A_0$--$A_1$ edge.  The same description applies to every deleted
edge inside $K$.  In either candidate world, each such edge has a Byzantine
endpoint.

Construct a mixed execution on $K$.  Keep the observations of $S$ as the only
external hypothesis-dependent inputs.  Give every process in $A_0\cap K$ an
auxiliary observation drawn from its hypothesis-$1$ law and every process in
$A_1\cap K$ one drawn from its hypothesis-$0$ law.  Fix canonical silence on
all edges missing from the pair graph.  In world~$0$, the $A_0$ processes are
Byzantine and simulate their auxiliary world-$1$ roles, while the $A_1$
processes are genuinely correct with exactly the auxiliary world-$0$ law.
World~$1$ is symmetric.  Because every controlled edge has a Byzantine endpoint
and a Byzantine process may sign arbitrary payloads under its own identity,
the same mixed evolution can be realized in both executions.

Thus for every finite-round protocol $\Pi_T$ there is a single stochastic
kernel $W_T$ with
\[
  \View_v^{0,F_0,\cA_0}(\Pi_T)
  \overset d= W_T(X_{S,0}^T,Z),
  \qquad
  \View_v^{1,F_1,\cA_1}(\Pi_T)
  \overset d= W_T(X_{S,1}^T,Z),
\]
where $Z$ has one hypothesis-independent law and is independent of the source
experiment $X_S^T$.  It includes correct-process random tapes, any
signature-generation coins, and the private randomization used by the simulated
Byzantine roles.  Randomized adversaries are admissible in our model; under a
deterministic-only convention, fixing an adversarial seed with error at least
its randomized average preserves the minimax lower bound
(Lemma~\ref{lem:deterministic-adversary}).  Conditional independence of source
observations is used precisely to obtain this common auxiliary variable.  The
target view is
therefore a post-processing of the centralized product experiment on $S$, so
Chernoff's theorem and data processing give
\[
  \mathcal E_v(\boldsymbol\Pi)
  \le \mathcal C_S.
\]
Minimizing over $F_0,F_1$ yields $\mathcal E_v^*\le\Lambda_v$.

\subsection{Achievability: cross-world causal-path certification}
A literal exposure proof could instantiate one signed audit for every simple
path.  That is unnecessary.  The achieving protocol is a single
full-information causal flood.

Process $i$ first signs its observation object $O_i$.  Its round-$0$ state
$C_i^0$ contains $O_i$.  In round $r\in\{1,\ldots,n-1\}$, correct $i$ sends
$C_i^{r-1}$ to every neighbor.  It then signs a new state $C_i^r$ containing its
previous self-state and, for each neighbor $j$, the unique valid round-$(r-1)$
state received from $j$ in the designated slot, or $\bot$ if none was received.
The target keeps $C_v^{n-1}$.  There is one transmission per oriented edge per
round, hence at most $2(n-1)|E|$ correct-process transmissions.

\paragraph{Why the next lemma is the conceptual step.}
Ordinary path certification takes place inside one execution: one argues that
all edges of a path were simultaneously between correct endpoints.  That is
\emph{false} here.  A path in $H(F_0,F_1)$ can contain edges whose correctness
is certified only in world~0 and other edges whose correctness is certified
only in world~1.  There need not be a single candidate world in which the
entire path is correct.  The reason the proof can nevertheless splice the path
is that it never splices hypothetical messages.  It moves backward through
exact immutable signed state objects that are already nested in the one target
transcript compatible with both worlds.

\begin{lemma}[Cross-world causal-path certification]
\label{lem:causal-path-exposure}
Fix two fault sets $F_0,F_1$ and a simple path
\[
    P=(u_0,u_1,\ldots,u_\ell=v)
\]
contained in $H(F_0,F_1)$, with $u_0\notin F_0\cup F_1$.  If one target
transcript $\tau_v$ is compatible with both $(0,F_0)$ and $(1,F_1)$, then the
final state $C_v^{n-1}$ contains a nested authenticated chain ending at
$C_{u_0}^0$, and hence contains the authentic observation certificate $O_{u_0}$.
\end{lemma}

\begin{proof}
Because $\ell\le n-1$, the self-history nested in $C_v^{n-1}$ contains the
concrete signed object $C_v^\ell$.  We work backward and maintain the following
invariant for $j=\ell,\ell-1,\ldots,0$:
\begin{quote}
\emph{$\mathcal I_j$: the concrete object $C_{u_j}^j$ has been exposed as a
subobject of $\tau_v$; moreover, in every compatible candidate world $h$ in
which $u_j\notin F_h$, this same object is the unique round-$j$ state generated
by correct process $u_j$.}
\end{quote}
The base case $\mathcal I_\ell$ holds because $u_\ell=v$ is correct in both
candidates and its round-$\ell$ self-state is part of the target history.

Assume $\mathcal I_j$ for some $j\ge1$.  Since
$\{u_{j-1},u_j\}$ is a pair-graph edge, choose a certifying world
$h_j\in\{0,1\}$ with $u_{j-1},u_j\notin F_{h_j}$.  By $\mathcal I_j$, the
exposed concrete object $C_{u_j}^j$ is exactly the state generated by correct
$u_j$ in that candidate execution.  Synchronous correct-to-correct delivery
therefore forces its designated receipt coordinate from $u_{j-1}$ to contain
the exact round-$(j-1)$ state sent by correct $u_{j-1}$, namely a concrete
object $C_{u_{j-1}}^{j-1}$.  Property (S4) binds that receipt coordinate and
its round tag inside the signed state, while (S2) permits the same concrete
object to be forwarded and nested unchanged.  Hence
$C_{u_{j-1}}^{j-1}$ is exposed in $\tau_v$.

It remains to establish the second clause of $\mathcal I_{j-1}$.  Consider any
compatible candidate world $h$ in which $u_{j-1}$ is correct.  By (S1) and the
fresh-session condition (S5), no other process and no pre-execution replay can
create a valid same-session object signed by $u_{j-1}$.  The protocol makes a
correct process sign exactly one state with tag $(u_{j-1},j-1)$.  Therefore the
exposed concrete object must be that unique correct state in candidate $h$.
This proves $\mathcal I_{j-1}$.

The induction also makes world switching explicit.  If consecutive path edges
are certified in different worlds, their shared vertex is correct in both
certifying worlds and hence lies outside $F_0\cup F_1$.  Equivalently, at a
vertex in $A_0=F_0\setminus F_1$ every retained incident edge can be certified
only in world~1, and at a vertex in $A_1=F_1\setminus F_0$ every retained
incident edge can be certified only in world~0.  Thus the certifying world can
switch only at a common-correct vertex, exactly where the same signed state is
forced to be authentic in both candidates.

At $j=0$ we obtain the nested chain
\[
C_v^\ell
\supset C_{u_{\ell-1}}^{\ell-1}
\supset\cdots\supset C_{u_0}^0
\supset O_{u_0}.
\]
Since $u_0\notin F_0\cup F_1$, the observation certificate is authentic in both
candidates.  No path is selected, tagged, or transmitted during communication;
the path is only a post-hoc proof/decoder witness.
\end{proof}

Thus every source in $S_v(F_0,F_1)$ is exposed to the pairwise decoder.  This
lemma, rather than full-information flooding by itself, is the mechanism that
matches the two-world converse without path enumeration.

For a realized transcript, consider every compatible candidate $(h,F)$.  For
each compatible opposite-hypothesis pair $(0,F_0),(1,F_1)$, the target can read
the authentic observations of $S_v(F_0,F_1)$ and form their log-likelihood
ratio.  Call a candidate a \emph{survivor} if it wins every direct comparison
against compatible candidates of the opposite hypothesis.  Opposite-hypothesis
survivors cannot coexist.  If $(0,F^\star)$ is the true candidate, an error
implies that it loses to at least one $F_1$.  For each fixed $F_1$,
\[
  \Pr_0[\text{loss to }F_1]
  \le
  e^{-T\mathcal C_{S_v(F^\star,F_1)}}
  \le e^{-T\Lambda_v}.
\]
The compatibility event need not be conditioned on: intersecting it with the
likelihood-loss event can only reduce probability.  A union bound over at most
$N_f=\sum_{j=0}^f\binom{n-1}{j}$ opposite fault sets gives
\[
  R_{v,T}\le \min\{1,N_f e^{-T\Lambda_v}\}.
\]
The hypothesis-$1$ case is symmetric.  Since $n$ and $f$ are fixed as
$T\to\infty$, the prefactor does not affect the exponent, giving
$\mathcal E_v^*\ge\Lambda_v$ and completing the theorem.

\section{Temporal Relational Observability}
\label{sec:temporal}

As a concrete application of Theorem~\ref{thm:main}, we instantiate the
hypothesis space with global event orders.  This section is not needed for the
pair-graph characterization.  Its combinatorial contribution asks how many
complete process traces must change before a globally coherent execution can
differ from the true execution by a prescribed amount.  For contiguous trace
windows, the answer has an exact cut form.  The main information theorem allows
fully heterogeneous process channels; the homogeneous trace-change assumption
appears only in the final closed-form statistical corollary below.

\subsection{Interval traces and relational distance}
\label{subsec:interval-model}

Let
\[
    \mathcal{E}=\{e_1,\ldots,e_m\}
\]
be a fixed set of events, and let the true global order be
\begin{equation}
    \label{eq:true-temporal-order}
    \pi=(e_1,e_2,\ldots,e_m).
\end{equation}
A set $\mathcal{R}\subseteq V\setminus\{v\}$ of processes acts as
\emph{reporters}; the target $v$ is an external, nonobserving monitor in this
section.  Reporter $i\in\mathcal{R}$ observes the contiguous event set
\begin{equation}
    \label{eq:reporter-interval}
    I_i=[\ell_i,r_i]
    :=\{e_{\ell_i},e_{\ell_i+1},\ldots,e_{r_i}\},
    \qquad
    1\le \ell_i\le r_i\le m.
\end{equation}
For any total order $\tau$ of $\mathcal{E}$, let $L_i(\tau)$ denote the
restriction of $\tau$ to the events in $I_i$.  Thus a coherent alternative
execution changes reporter $i$ precisely when
\[
    L_i(\tau)\neq L_i(\pi),
\]
while preserving reporter $i$'s event inventory exactly.

For another total order $\sigma$, define its Kendall distortion from $\pi$ by
\begin{equation}
    \label{eq:kendall-distortion}
    K(\pi,\sigma)
    :=
    \left|
        \left\{
        (a,b):1\le a<b\le m,\ e_b\prec_{\sigma} e_a
        \right\}
    \right|.
\end{equation}
Let
\begin{equation}
    \label{eq:changed-reporters}
    D(\pi,\sigma)
    :=
    \{i\in\mathcal{R}:L_i(\sigma)\neq L_i(\pi)\}
\end{equation}
be the reporters whose complete local traces change.

\begin{definition}[Temporal relational distance]
\label{def:temporal-drel}
For an integer $0\le \Gamma\le \binom{m}{2}$, define
\begin{equation}
    \label{eq:temporal-drel}
    d_{\mathrm{rel}}^{\pi}(\Gamma)
    :=
    \min_{\substack{\sigma\text{ a total order of }\mathcal{E}\\
                     K(\pi,\sigma)\ge \Gamma}}
    |D(\pi,\sigma)|.
\end{equation}
Thus $d_{\mathrm{rel}}^{\pi}(\Gamma)$ is the minimum number of reporter traces
that a globally coherent alternative must change in order to move at least
$\Gamma$ pairwise temporal relations.
\end{definition}

The definition in \eqref{eq:temporal-drel} appears to require a minimization over
$m!$ global orders.  Contiguous observation intervals collapse that search to a
cut problem on the true timeline.

\subsection{Boundary normal form}
\label{subsec:boundary-normal-form}

For every adjacent boundary $k\in[m-1]$, define its witness set
\begin{equation}
    \label{eq:boundary-witness-set}
    C_k
    :=
    \{i\in\mathcal{R}:\ell_i\le k<r_i\}.
\end{equation}
These are exactly the reporters that observe both $e_k$ and $e_{k+1}$.

For a total order $\sigma$, define its set of reversed true-order boundaries by
\begin{equation}
    \label{eq:boundary-set-sigma}
    B(\sigma)
    :=
    \{k\in[m-1]: e_{k+1}\prec_{\sigma}e_k\}.
\end{equation}

\begin{lemma}[Changed reporters are boundary witnesses]
\label{lem:changed-equals-boundary-union}
For every total order $\sigma$,
\begin{equation}
    \label{eq:changed-equals-boundary-union}
    D(\pi,\sigma)
    =
    \bigcup_{k\in B(\sigma)} C_k.
\end{equation}
\end{lemma}

\begin{proof}
If $i\in C_k$ for a reversed boundary $k$, then $i$ observes an adjacent pair
whose order changed, so $i\in D(\pi,\sigma)$.  Conversely, if $i$ witnesses no
reversed internal boundary, every adjacent pair in its contiguous interval
retains increasing order; transitivity gives $L_i(\sigma)=L_i(\pi)$.  This
proves both inclusions.
\end{proof}

For a boundary set $B\subseteq[m-1]$, delete the boundaries in $B$ from the
chain
\[
    e_1-e_2-\cdots-e_m.
\]
This partitions the true order into maximal contiguous blocks
\[
    J_1,J_2,\ldots,J_q
\]
with sizes $s_1,\ldots,s_q$.  Define
\begin{align}
    \label{eq:temporal-cut-cost}
    \operatorname{cost}(B)
    &:=
    \left|\bigcup_{k\in B}C_k\right|,\\
    \label{eq:temporal-cut-cap}
    \operatorname{cap}(B)
    &:=
    \sum_{1\le a<b\le q}s_as_b
      =
    \binom{m}{2}-\sum_{j=1}^{q}\binom{s_j}{2}.
\end{align}
The quantity $\operatorname{cap}(B)$ counts the unordered event pairs that lie
in different blocks.

\begin{lemma}[Boundary capacity upper-bounds distortion]
\label{lem:boundary-capacity-upper}
For every total order $\sigma$,
\begin{equation}
    \label{eq:kendall-cap-upper}
    K(\pi,\sigma)
    \le
    \operatorname{cap}(B(\sigma)).
\end{equation}
\end{lemma}

\begin{proof}
Let $J_1,\ldots,J_q$ be the blocks induced by $B(\sigma)$.  If two consecutive
true-order events belong to the same block, their boundary is not in
$B(\sigma)$ and therefore their relative order is preserved by $\sigma$.
Transitivity implies that every block $J_j$ occurs internally in the same order
under $\sigma$ as under $\pi$.  Consequently, no pair of events contained in a
single block contributes a Kendall inversion.  Every inversion must cross two
distinct blocks, and there are exactly $\operatorname{cap}(B(\sigma))$ such
pairs.
\end{proof}

\begin{lemma}[Canonical block-reversal witness]
\label{lem:block-reversal-witness}
For every $B\subseteq[m-1]$, there exists a total order $\sigma_B$ satisfying
\begin{equation}
    \label{eq:block-reversal-properties}
    B(\sigma_B)=B,
    \qquad
    K(\pi,\sigma_B)=\operatorname{cap}(B).
\end{equation}
\end{lemma}

\begin{proof}
List the induced blocks $J_1,\ldots,J_q$ from left to right and set
$\sigma_B=(J_q,\ldots,J_1)$, preserving each block internally.  Exactly the
selected boundaries reverse, every within-block pair is preserved, and every
cross-block pair reverses; hence $B(\sigma_B)=B$ and
$K(\pi,\sigma_B)=\operatorname{cap}(B)$.
\end{proof}

Lemmas~\ref{lem:changed-equals-boundary-union}--\ref{lem:block-reversal-witness}
remove permutations from the optimization entirely.

\begin{theorem}[Exact temporal-cut characterization]
\label{thm:temporal-cut}
For every $0\le \Gamma\le\binom{m}{2}$,
\begin{equation}
    \label{eq:temporal-cut-theorem}
    \boxed{
    d_{\mathrm{rel}}^{\pi}(\Gamma)
    =
    \min_{\substack{B\subseteq[m-1]\\
                     \operatorname{cap}(B)\ge\Gamma}}
    \operatorname{cost}(B)
    =
    \min_{\substack{B\subseteq[m-1]\\
                     \operatorname{cap}(B)\ge\Gamma}}
    \left|\bigcup_{k\in B}C_k\right|.
    }
\end{equation}
\end{theorem}

\begin{proof}
For any feasible $\sigma$, set $B=B(\sigma)$.  Lemmas~\ref{lem:boundary-capacity-upper}
and~\ref{lem:changed-equals-boundary-union} give
$\operatorname{cap}(B)\ge K(\pi,\sigma)\ge\Gamma$ and
$|D(\pi,\sigma)|=\operatorname{cost}(B)$, proving the lower bound.  Conversely,
for any $B$ with $\operatorname{cap}(B)\ge\Gamma$, the block-reversal order
$\sigma_B$ has exactly that distortion and reporter cost.  Minimizing proves
equality.
\end{proof}

A useful check is $d_{\mathrm{rel}}^{\pi}(1)=\min_k|C_k|$.  More generally, a
single boundary $k$ has capacity $k(m-k)$, so the theorem trades reporter-union
cost against cross-block inversion capacity rather than merely counting cuts.

\subsection{From temporal distance to an exact Byzantine error exponent}
\label{subsec:temporal-statistical}

We next connect Theorem~\ref{thm:temporal-cut} to the pair-graph
characterization of Sections~\ref{sec:pair-graph}--\ref{sec:proof-main}.  The
clean closed form requires two explicit conditions: the network must be
$f$-pair-connected at the target, and every changed reporter must contribute
the same Chernoff profile.

Recall Definition~\ref{def:pair-connected-target}.  Under $f$-pair-connectivity at $v$,
\begin{equation}
    \label{eq:Sv-pair-connected}
    S_v(F_0,F_1)=V\setminus(F_0\cup F_1).
\end{equation}
Corollary~\ref{cor:pair-connected-general} shows that, in this regime, the
network contribution reduces exactly to the deletion of at most $2f$ sources.

We now specify the statistical specialization.  For every possible local trace
$\ell$ of reporter $i$, let $Q_{i,\ell}$ be its per-sample observation law.  If
the global order is $\tau$, reporter $i$ obtains $T$ conditionally independent
samples from $Q_{i,L_i(\tau)}$.  Nonreporting relay processes and the external
target have hypothesis-independent observations.

Fix the baseline order $\pi$.  Assume that there is one function
$\chi:[0,1]\to\mathbb{R}_{\ge0}$ such that for every reporter $i$ and every local
trace $\ell\neq L_i(\pi)$,
\begin{equation}
    \label{eq:common-chernoff-profile}
    -\log
    \sum_x
    q_{i,L_i(\pi)}(x)^{1-s}q_{i,\ell}(x)^s
    =
    \chi(s)
    \qquad
    \text{for all }s\in[0,1].
\end{equation}
Let
\begin{equation}
    \label{eq:homogeneous-c}
    c:=\max_{s\in[0,1]}\chi(s).
\end{equation}
We call \eqref{eq:common-chernoff-profile} the \emph{homogeneous trace-change
model around $\pi$}.  It is stronger than merely requiring every changed trace
to have the same scalar Chernoff information: equality of the whole affinity
profile ensures exact additivity for heterogeneous changed traces under one
common Chernoff tilt.

\paragraph{Concrete channel family.}
The homogeneity condition is nonempty.  For fixed $p\ne q$, take
$Q_{i,L_i(\pi)}=\operatorname{Bern}(p)$ for every reporter and
$Q_{i,\ell}=\operatorname{Bern}(q)$ for every changed trace.  Then
$\chi(s)=-\log\!\bigl(p^{1-s}q^s+(1-p)^{1-s}(1-q)^s\bigr)$ is independent of
both the reporter and the changed trace.

\begin{corollary}[Temporal specialization of the pair theorem]
\label{cor:temporal-exponent}
Assume $f$-pair-connectivity at an external (or hypothesis-independent) target
and the homogeneous trace-change profile~\eqref{eq:common-chernoff-profile}.
Then every alternative order $\sigma$ satisfies
\begin{equation}
    \label{eq:pairwise-temporal-exponent}
    \mathcal E_v^\star(\pi,\sigma)
    =c[|D(\pi,\sigma)|-2f]_+.
\end{equation}
Moreover, for $1\le\Gamma\le\binom m2$, if
$\mathcal A_\Gamma(\pi)=\{\sigma\ne\pi:K(\pi,\sigma)\ge\Gamma\}$, then the
optimal minimax exponent for testing $\pi$ against this composite alternative is
\begin{equation}
    \label{eq:temporal-exponent}
    \boxed{
    \mathcal E^\star_{v,\Gamma}(\pi)
    =c[d_{\rm rel}^{\pi}(\Gamma)-2f]_+ .}
\end{equation}
If $c>0$, a positive exponent is possible exactly when
$d_{\rm rel}^{\pi}(\Gamma)>2f$.  Without this nondegeneracy condition,
formula~\eqref{eq:temporal-exponent} remains valid, but the exponent may be zero
even when the relational distance exceeds $2f$.
\end{corollary}

\begin{proof}
Under~\eqref{eq:common-chernoff-profile}, unchanged reporters contribute zero
and every changed reporter contributes the same affinity profile, so
$\mathcal C_S^{\pi,\sigma}=c|S\cap D(\pi,\sigma)|$.  Pair-connectivity gives
$S_v(F_0,F_1)=V\setminus(F_0\cup F_1)$, and Theorem~\ref{thm:main} therefore
allows the two fault sets to remove exactly $\min\{2f,|D(\pi,\sigma)|\}$ changed
reporters, proving the pairwise formula.  The family $\mathcal A_\Gamma(\pi)$ is
fixed and finite as $T\to\infty$; the same pairwise decoder and a finite union
bound make its exponent the minimum baseline-versus-alternative exponent.  No
distinction among alternative orders is required.  Theorem~\ref{thm:temporal-cut}
then gives~\eqref{eq:temporal-exponent}; Appendix~\ref{app:multi} formalizes
the distinction between finite-hypothesis identification and this
baseline-versus-composite test.
\end{proof}

\paragraph{Scope of the specialization.}
Homogeneity is used only in this corollary.  The main pair-graph theorem permits
heterogeneous process channels; nonhomogeneous temporal profiles simply replace
the reporter count by the corresponding weighted product Chernoff optimization.

\paragraph{Secondary consequences.}
Appendix~\ref{app:sparse} shows that $2f$-vertex-fault-tolerant spanners preserve
all pair components and hence $\Lambda_v$ while reducing transmission count;
Appendix~\ref{app:limits} collects the precise model assumptions and scope.

\section{Related Work}
\label{sec:related}

Our theorem connects two mature literatures through a graph-dependent residual
\emph{statistical experiment}.  Authentication and topology determine
$S_v(F_0,F_1)$, and the same experiment gives both an all-protocol converse and
a matching authenticated causal certificate.  Table~\ref{tab:related-comparison} summarizes the main
separation.

\begin{table}[t]
\centering
\caption{Closest lines of work and the quantity they characterize.  The point
of comparison is the output of the theory, not merely the presence of
Byzantine faults or Chernoff exponents.}
\label{tab:related-comparison}
\small
\begin{tabularx}{\textwidth}{@{}>{\raggedright\arraybackslash}p{0.24\textwidth}>{\raggedright\arraybackslash}p{0.30\textwidth}>{\raggedright\arraybackslash}X@{}}
\toprule
Line of work & What is characterized & Distinction from this paper \\
\midrule
Centralized Byzantine inference
\cite{KosutTong2008Source,KosutTong2008CEO,MaranoMattaTong2009,RenYanMo2018}
& Source-coding feasibility, fake-distribution ambiguity, or detection error-exponent/security tradeoffs under compromised sensors
& Establishes the statistical two-world and $2f$-type phenomena in centralized or fusion-center settings; this paper characterizes the residual experiment selected by an authenticated multi-hop graph. \\
\addlinespace
Authenticated communication
\cite{DolevStrong1983,BeimelFranklin1999,Pagourtzis2017,ChotkanEtAl2024}
& Whether a particular value can be delivered or agreed upon under Byzantine faults, using topology, paths, cuts, signatures, or trust
& Focuses on authentic delivery or agreement of particular values; this paper assigns heterogeneous statistical value to source identity and optimizes the testing exponent over all finite-round interactive protocols. \\
\addlinespace
This work
& The exact product experiment a fixed target must retain, and the resulting minimax error exponent over all finite-round authenticated interaction
& Couples two fault explanations, topology, target location, and source informativeness; cross-world certification matches the converse. \\
\bottomrule
\end{tabularx}
\end{table}

\paragraph{Centralized Byzantine inference.}
Kosut and Tong's Byzantine source-coding converse is a particularly close
conceptual precursor~\cite{KosutTong2008Source}: traitors simulate a fake source
distribution so that several honest/traitor explanations remain compatible
with a fusion decoder's observations.  Their Byzantine CEO work studies an
error exponent under analogous ambiguity~\cite{KosutTong2008CEO}.  Marano,
Matta, and Tong optimize fictitious Byzantine reports to degrade detection at a
fusion center~\cite{MaranoMattaTong2009}, while Ren, Yan, and Mo characterize
fundamental security/efficiency tradeoffs for binary testing with compromised
sensors~\cite{RenYanMo2018}.  Thus Byzantine hypothesis testing, Chernoff
exponents, fake statistical worlds, alternative plausible traitor sets, and
$2f$-style redundancy are established ideas.

Our theorem recovers that phenomenon on pair-connected graphs:
Corollary~\ref{cor:pair-connected-general} gives
$\Lambda_v=\min_{|U|\le2f}\mathcal C_{V\setminus U}$.  The new issue is what
replaces this expression on an incomplete authenticated graph.  Two candidate
fault explanations can hand a simulation across an edge only when opposite
endpoints are faulty in opposite worlds.  Removing exactly those handoffs
produces $K_v(F_0,F_1)$ and hence the residual product experiment
$S_v(F_0,F_1)$.  Example~\ref{ex:weighted-same-topology} shows why a graph-only
resilience parameter is insufficient: the communication topology is unchanged
while moving the informative sources changes the optimal exponent from $9$ to
$2$.

\paragraph{Authenticated communication and general adversaries.}
Dolev--Strong established authenticated synchronous agreement
\cite{DolevStrong1983}; later work studies authenticated consensus on arbitrary
or directed graphs~\cite{ChlebusKowalskiOlkowski2020,VaidyaTseng2026} and
reliable communication or broadcast under partial authentication, incomplete
topology, or hybrid trust
\cite{BeimelFranklin1999,Pagourtzis2017,BonomiEtAl2024,ChotkanEtAl2024}.
These results ask whether a particular payload can be transmitted or agreed
upon.  General-adversary simulation studies corruption structures
\cite{HirtMaurer2000}, and Byzantine network coding studies validity and
throughput under network adversaries~\cite{KosutTongTse2010}.  Our objective is
different: an individual source can be undeliverable while substitutable
statistical evidence still yields a positive inference exponent.  The pair
graph therefore asks not whether every source can communicate reliably, but
which source experiment must remain distinguishable under two plausible fault
explanations.

\paragraph{Distributed Byzantine inference.}
Other work studies Bayesian detection under Byzantine data, resilient
hypothesis learning, and time-varying networks
\cite{Kailkhura2015,MitraRichardsSundaram2019,WuEtAl2022}.  These results
provide important robust detectors and learning rules.  The present theorem
instead optimizes over the entire class of finite-round authenticated
interactive protocols and characterizes the optimum by a target-specific graph
object.  The flood itself is standard.  The new argument is the sufficiency
proof: one signed causal history exposes exactly the sources selected by the
pair graph even when the certifying candidate world changes along a path.

\paragraph{Orders, Kendall distance, and distributed agreement.}
Wood, Mendes, and Pulaj study preference aggregation with crash faults,
relaxing consensus to $k$-set or $\varepsilon$-approximate agreement on
rankings; their impossibility results include Kendall-tau and
Spearman-footrule metrics~\cite{WoodMendesPulaj2024}.  Their objective is
agreement on an output ranking from process preferences.  Our temporal section
instead posits a latent event order, partial interval traces, and Byzantine
communication, and asks how much statistical evidence a fixed target retains.
The boundary-cut theorem measures how many reporter traces must change to
produce a prescribed Kendall distortion; Theorem~\ref{thm:main} then converts
that combinatorial quantity into a target-specific inference exponent.
Remark~\ref{rem:target-vs-agreement} separates this objective explicitly from
agreement-constrained inference.

\section{Conclusion}
\label{sec:conclusion}
Authentication does not by itself determine which statistical evidence survives
Byzantine communication.  The authenticated pair graph gives the exact answer
for the model studied here: for every two fault explanations it identifies the
common-correct source experiment that cannot be hidden from the target, and the
minimum Chernoff information of those experiments is exactly the optimal
minimax exponent over all finite-round interactive protocols.  The converse is
a two-world simulation; the matching direction is the cross-world
causal-certification lemma showing that one standard signed causal history
contains the required evidence even when different path edges are justified in
different candidate worlds.

The temporal boundary theorem illustrates how the abstract information
parameter can collapse to a concrete combinatorial quantity.  Sparse
preservation and the W[1]-hardness result describe, respectively, a structural
robustness property and a computational boundary of the characterization.
A natural next question is whether the exact exponent can also be attained
with succinct bit complexity and efficient decoding under natural
cryptographic instantiations.

\bibliography{arxiv_references}

\appendix

\section{Full Proof of the Exact Pair-Graph Characterization}
\label{app:full-proof}

We prove Theorem~\ref{thm:main} in two steps.  Section~\ref{subsec:converse}
shows that no interactive protocol can beat $\Lambda_v$.  Section~\ref{subsec:achievability}
constructs a finite-round authenticated protocol attaining $\Lambda_v$.

\subsection{Converse: a universal two-world simulation}
\label{subsec:converse}

Fix arbitrary candidate fault sets $F_0,F_1$ of size at most $f$, and abbreviate
\[
    B=F_0\cap F_1,
    \qquad
    A_0=F_0\setminus F_1,
    \qquad
    A_1=F_1\setminus F_0,
\]
\[
    H=H(F_0,F_1),
    \qquad
    K=K_v(F_0,F_1),
    \qquad
    S=S_v(F_0,F_1).
\]

The first lemma isolates the target's pair component from the rest of the graph
in both candidate worlds.

\begin{lemma}[Controlled boundary]
\label{lem:controlled-boundary}
Let $x\in K$ and $y\notin K$ with $\{x,y\}\in E$.  Then at least one endpoint
of $\{x,y\}$ is Byzantine under fault set $F_0$, and at least one endpoint is
Byzantine under fault set $F_1$.
\end{lemma}

\begin{proof}
Because $x\in K$ and $y\notin K$, the edge $\{x,y\}$ is absent from the pair
graph $H$.  By Definition~\ref{def:pair-graph}, this can happen only if either
$y\in B$ (a vertex in $B$ is deleted altogether) or the edge joins a process in
$A_0$ to a process in $A_1$.

If $y\in B$, then $y\in F_0\cap F_1$ and the claim is immediate.  Otherwise the
edge is an $A_0$--$A_1$ edge.  In world $0$, the endpoint in $A_0$ belongs to
$F_0$; in world $1$, the endpoint in $A_1$ belongs to $F_1$.  Hence every edge
crossing $K$ has a Byzantine endpoint in either candidate world.
\end{proof}

Call an original edge \emph{controlled} if it has exactly one endpoint in $K$,
or if it is an internal $A_0$--$A_1$ edge deleted from $H[K]$.  By
Lemma~\ref{lem:controlled-boundary} and the definition of $A_0,A_1$, every
controlled edge has a Byzantine endpoint in each candidate world.  Hence the
adversary can impose the prescribed canonical input at every controlled
interface.  The execution restricted to $K$ is therefore closed: every input
delivered to a correct process in $K$ is determined either by an $H[K]$ edge
inside $K$ or by canonical silence on a controlled edge, so no process outside
$K$ can inject uncontrolled information into the target view.

\begin{lemma}[Two-world simulation]
\label{lem:two-world-simulation}
Fix any finite-round randomized authenticated protocol $\Pi_T$.  There exist a
stochastic kernel $W_T$ and Byzantine strategies $\cA_0,\cA_1$ for the two
executions
\[
    (\theta,F)=(0,F_0)
    \qquad\text{and}\qquad
    (\theta,F)=(1,F_1)
\]
such that
\begin{align}
    \label{eq:view-kernel-world0}
    \View_v^{0,F_0,\cA_0}(\Pi_T)
    &\overset{d}{=}
    W_T\bigl(X_{S,0}^T,Z\bigr),\\
    \label{eq:view-kernel-world1}
    \View_v^{1,F_1,\cA_1}(\Pi_T)
    &\overset{d}{=}
    W_T\bigl(X_{S,1}^T,Z\bigr),
\end{align}
where $X_{S,h}^T\sim\bigotimes_{i\in S}P_{i,h}^{\otimes T}$ and $Z$ has the
same distribution in the two equations and is independent of $X_{S,h}^T$.
Thus the only hypothesis-dependent inputs to the common channel are the
observations of the processes in $S$.
\end{lemma}

\begin{proof}
We construct an auxiliary mixed execution on the pair-graph component $K$.

For each $i\in S$, leave $i$'s observation block as an external input $x_i^T$.
For each $i\in A_0\cap K$, sample an auxiliary block
\[
    \widetilde X_i^T\sim P_{i,1}^{\otimes T},
\]
and for each $i\in A_1\cap K$, sample
\[
    \widetilde X_i^T\sim P_{i,0}^{\otimes T}.
\]
Sample independent private random tapes for every simulated correct role.  All
messages entering a correct process of $K$ through an edge absent from $H[K]$
are fixed to canonical silence.  Running the correct protocol transition rules
inside $H[K]$ with these inputs defines a stochastic kernel from $x_S^T$ to the
target view.  Let $Z$ collect all auxiliary observation blocks, all private
random tapes, and all other hypothesis-independent randomization used by this
construction; call the resulting kernel $W_T$.

We now realize this same mixed execution inside the two actual Byzantine worlds.

\paragraph{World 0.}
Let the true hypothesis be $0$ and the Byzantine set be $F_0$.  Every process in
$S$ is correct and has its genuine $P_{i,0}^{\otimes T}$ observation block.
Every process in $A_1\cap K$ is also correct in world $0$, and its genuine
observation law is exactly the auxiliary law $P_{i,0}^{\otimes T}$ assigned in
the mixed execution.  Processes in $A_0\cap K$ are Byzantine.  The adversary
samples the auxiliary $P_{i,1}^{\otimes T}$ blocks and random tapes assigned to
those roles and makes each Byzantine process simulate the outgoing behavior of
that correct mixed-execution role.  Since a Byzantine process owns its own
signing identity, every signature required by its simulated role can be produced
legally.

Every edge through which the mixed execution prescribes canonical silence but
which is present in the original graph has a Byzantine endpoint in world $0$:
this follows from Lemma~\ref{lem:controlled-boundary} for edges leaving $K$, and
from the definition of $E(A_0,A_1)$ for deleted internal edges.  The adversary
therefore enforces the prescribed silent input at the correct endpoint and
ignores any message delivered to a faulty endpoint.

\paragraph{World 1.}
The construction is symmetric.  Processes in $S$ are correct with genuine
$P_{i,1}^{\otimes T}$ observations; processes in $A_0\cap K$ are correct with
genuine $P_{i,1}^{\otimes T}$ observations; processes in $A_1\cap K$ are
Byzantine and simulate the auxiliary $P_{i,0}^{\otimes T}$ roles.  Every edge
removed from $H[K]$ again has a Byzantine endpoint and can be made to present the
same canonical silence to the correct endpoint.

It remains to justify that randomized protocols and rushing do not invalidate
the coupling.  We prove by induction on synchronous rounds that, conditioned on
the sampled variables collected in $Z$, the local states of all simulated roles
in $K$ are identical to their states in the mixed execution.  This is immediate
before round $1$.  Suppose it holds before round $r$.  Correct round-$r$ outgoing
messages are then determined by the corresponding mixed states and therefore
match the mixed execution.  Byzantine simulators may rush, observe all correct
round-$r$ outgoing messages, and choose their own messages to match their mixed
roles.  The controlled edges present the prescribed canonical silence, so all
correct roles receive exactly the mixed round-$r$ inputs and the induction
continues.

The target $v$ belongs to $S$, hence is correct in both worlds and follows the
same mixed transition rule.  Equations~\eqref{eq:view-kernel-world0} and
\eqref{eq:view-kernel-world1} follow.  Conditional independence
\eqref{eq:product-observation-law} ensures that the auxiliary observations
included in $Z$ are independent of $X_S^T$ and can have the same law in the two
channel representations.  Any private coins used to generate signatures are
included among the hypothesis-independent random tapes in $Z$; conditioned on
those coins, the proof couples the concrete signed objects produced in the
mixed execution.
\end{proof}

\begin{lemma}[Randomized versus deterministic Byzantine strategies]
\label{lem:deterministic-adversary}
The converse remains valid if the admissible Byzantine strategies in
\eqref{eq:minimax-risk} are required to be deterministic.
\end{lemma}
\begin{proof}
The construction in Lemma~\ref{lem:two-world-simulation} may use a private
adversarial seed $\rho_h$ in world $h$.  For fixed $T$, let
$e_h(\rho_h)$ be the target's error probability after fixing that seed, where
the remaining probability is over source observations and correct-process
randomness.  The error $e_h$ of the randomized strategy is
$\mathbb E_{\rho_h}[e_h(\rho_h)]$, so some deterministic seed
$\rho_h^\star$ satisfies $e_h(\rho_h^\star)\ge e_h$.  The minimax risk takes a
separate supremum over the Byzantine strategy for each hypothesis and fault
set.  We may therefore fix $\rho_0^\star$ and $\rho_1^\star$ separately, and
its lower bound by $(e_0+e_1)/2$ is unchanged.  Randomization is thus a
convenient way to describe the common-channel simulation, not an extra source
of power needed for the converse.
\end{proof}

\begin{proposition}[Converse]
\label{prop:converse}
Every protocol family $\boldsymbol\Pi$ satisfies
\[
    \cE_v(\boldsymbol\Pi)\le\Lambda_v.
\]
Consequently,
\[
    \cE_v^\star\le\Lambda_v.
\]
\end{proposition}

\begin{proof}
Fix $F_0,F_1$ and $S=S_v(F_0,F_1)$.  Apply
Lemma~\ref{lem:two-world-simulation} to $\Pi_T$.  Under the two adversarial
executions constructed there, the target view is obtained by passing the
centralized observations $X_S^T$ through the same stochastic channel $W_T$.
Therefore any decision rule based on the target view can be simulated by a
(possibly randomized) centralized test that first applies $W_T$ to $X_S^T$ and
then runs the target decision rule.  Such a test cannot have smaller equal-prior
Bayes error than the optimal centralized test that observes $X_S^T$ directly.

Let $e_0(T)$ and $e_1(T)$ be the target's error probabilities in the two
constructed executions.  Since the minimax risk dominates their average,
\[
    R_{v,T}(\Pi_T)
    \ge
    \frac{e_0(T)+e_1(T)}{2}
    \ge
    P_{e,S}^\star(T).
\]
By \eqref{eq:chernoff-theorem},
\[
    \limsup_{T\to\infty}
    -\frac{1}{T}\log R_{v,T}(\Pi_T)
    \le
    \cC_S.
\]
The pair $(F_0,F_1)$ was arbitrary, hence
\[
    \cE_v(\boldsymbol\Pi)
    \le
    \min_{F_0,F_1}\cC_{S_v(F_0,F_1)}
    =
    \Lambda_v.
\]
Taking the supremum over protocol families proves the proposition.
\end{proof}

\subsection{Achievability: signed causal-state flooding}
\label{subsec:achievability}

We now give the protocol used in Theorem~\ref{thm:main}.  It is a
full-information protocol at the level of payload contents, but it schedules
only one state transmission per neighbor per round.

\subsubsection{The causal-state protocol}

Before communication begins, every process $i$ forms an observation certificate
\begin{equation}
    \label{eq:observation-cert}
    O_i
    :=
    [\sid,\mathsf{obs},i,X_i^T]_i.
\end{equation}
Its initial causal state is
\begin{equation}
    \label{eq:causal-state0}
    C_i^0
    :=
    [\sid,\mathsf{state},i,0,O_i]_i.
\end{equation}

Set $R=n-1$.  At the beginning of each round $r\in\{1,\ldots,R\}$, every
correct process $i$ sends the same state $C_i^{r-1}$ to every neighbor.  At the
end of the round, for each neighbor $j\in N(i)$, let $Y_{i\leftarrow j}^r$ be
the unique well-formed, correctly tagged state certificate signed by $j$ for
round $r-1$ that was delivered in the designated protocol slot; if no such
unique state was delivered, set $Y_{i\leftarrow j}^r=\botmsg$.  A state is
well formed recursively: its signer, session and round tags are correct and all
nested state objects have the prescribed syntax and valid signatures.

Correct $i$ then forms
\begin{equation}
    \label{eq:causal-state-r}
    C_i^r
    :=
    [\sid,\mathsf{state},i,r,C_i^{r-1},
      ((j,Y_{i\leftarrow j}^r))_{j\in N(i)}]_i.
\end{equation}

The target includes its final state $C_v^R$ in its transcript.  Because each
state embeds the preceding self-state, $C_v^R$ also contains $C_v^r$ for every
$r\le R$.

The protocol schedules one state transmission per oriented edge per round.
Thus correct processes send at most
\begin{equation}
    \label{eq:causal-message-count}
    2R|E|=2(n-1)|E|
\end{equation}
correct-process transmissions.  Byzantine processes may choose arbitrary payloads in their protocol
slots; malformed or multiply supplied states are treated as $\botmsg$.

\subsubsection{Compatibility}

\begin{definition}[Candidate compatibility]
\label{def:compatibility}
A candidate pair $(h,F)$, where $h\in\{0,1\}$ and
$F\subseteq V\setminus\{v\}$ with $|F|\le f$, is \emph{compatible} with a target
transcript $\tau_v$ if there exist
\begin{enumerate}
\renewcommand{\labelenumi}{(\roman{enumi})}
    \item observation blocks for all processes outside $F$ that lie in the
          support of their distributions under hypothesis $h$,
    \item private random tapes for all correct processes, and
    \item a Byzantine strategy for the processes in $F$,
\end{enumerate}
such that the causal-state protocol produces target transcript $\tau_v$.
\end{definition}

The true pair $(\theta^\star,F^\star)$ is compatible with the realized transcript
by construction.  Compatibility is information-theoretic; the decoder need not
check it in polynomial time.

\begin{lemma}[Every pair-graph edge has a correct world]
\label{lem:pair-edge-correct-world}
If $\{x,y\}$ is an edge of $H(F_0,F_1)$, then there exists
$h\in\{0,1\}$ such that $x,y\notin F_h$.
\end{lemma}

\begin{proof}
Neither endpoint belongs to $B=F_0\cap F_1$.  Hence each endpoint is either
common-correct, in $A_0=F_0\setminus F_1$, or in
$A_1=F_1\setminus F_0$.  The pair graph deletes every $A_0$--$A_1$ edge.
Thus if an endpoint is in $A_0$, the other endpoint is either common-correct or
also in $A_0$, so both are correct in world~1.  Symmetrically, an edge incident
to $A_1$ but not $A_0$--$A_1$ has both endpoints correct in world~0.  Two
common-correct endpoints are correct in both worlds.
\end{proof}

\paragraph{Cross-world path certification.}
The key structural lemma is stated and proved in the main text as
Lemma~\ref{lem:causal-path-exposure}.  The preceding edge lemma isolates its
only graph-theoretic premise: every retained pair-graph edge has at least one
candidate world in which both endpoints are correct.  The main-text proof then
uses authentication to switch those certifying worlds edge by edge inside one
immutable target transcript.

\begin{corollary}[Exposure of the common-correct pair component]
\label{cor:component-exposure}
Let one transcript $\tau_v$ be compatible with $(0,F_0)$ and $(1,F_1)$.  Then
for every $i\in S_v(F_0,F_1)$, the transcript exposes the authentic observation
certificate $O_i$.
\end{corollary}

\begin{proof}
If $i=v$, the target already has $X_v^T$.  Otherwise, because
$i\in K_v(F_0,F_1)$, choose any simple $i$--$v$ path in $H(F_0,F_1)$ and apply
Lemma~\ref{lem:causal-path-exposure}.  No such path is selected or transmitted
during the communication phase; it is chosen only during local decoding.
\end{proof}

\subsubsection{Pairwise likelihood comparisons}

Fix a realized transcript $\tau_v$.  For every two compatible candidates
$(0,F_0)$ and $(1,F_1)$, Corollary~\ref{cor:component-exposure} allows $v$ to
recover the genuine observation blocks of all processes in
$S_v(F_0,F_1)$.  For each $i\in S_v(F_0,F_1)$, process $i$ is correct in both
candidates.  By (S1), (S5), and the protocol rule that a correct process creates
exactly one observation certificate in the session, compatibility cannot expose
two inconsistent valid $i$-certificates for this pair.  The observation block
used below is therefore unique, so the likelihood ratio is well defined.
Define the pairwise log-likelihood ratio
\begin{equation}
    \label{eq:pairwise-llr}
    \LLR(F_0,F_1)
    :=
    \sum_{i\in S_v(F_0,F_1)}
    \sum_{t=1}^{T}
    \log
    \frac{p_{i,0}(X_{i,t})}{p_{i,1}(X_{i,t})}.
\end{equation}
We say that candidate $(0,F_0)$ \emph{wins} its direct comparison with
$(1,F_1)$ if $\LLR(F_0,F_1)\ge0$; otherwise $(1,F_1)$ wins.  The weak inequality is
an arbitrary deterministic tie rule.

A compatible candidate $(h,F)$ is called a \emph{survivor} if it wins its direct
comparison against every compatible candidate of the opposite hypothesis.  A
hypothesis-$0$ survivor and a hypothesis-$1$ survivor cannot coexist, because
the direct comparison between them has exactly one winner.  The decoder outputs
the hypothesis of the survivors if any survivor exists and breaks ties
arbitrarily if there is no survivor.

\begin{lemma}[Finite-sample error bound]
\label{lem:finite-sample-ach}
Let
\[
    N_f:=\sum_{j=0}^{f}\binom{n-1}{j}.
\]
The causal-state protocol and survivor decoder satisfy
\begin{equation}
    \label{eq:finite-error-bound}
    R_{v,T}
    \le
    \min\left\{1,\; N_f e^{-T\Lambda_v}\right\}.
\end{equation}
\end{lemma}

\begin{proof}
Consider first an arbitrary actual execution with true hypothesis $0$, actual
fault set $F^\star$, and arbitrary Byzantine strategy.  The true candidate
$(0,F^\star)$ is compatible with the realized transcript.  If it wins every
direct comparison against compatible hypothesis-$1$ candidates, then it is a
survivor and the decoder outputs $0$.  Therefore an error implies that there
exists a fault set $F_1$ such that $(1,F_1)$ is compatible and
\[
    \LLR(F^\star,F_1)<0.
\]
For a fixed $F_1$, the event
\[
    \{(1,F_1)\text{ compatible and }\LLR(F^\star,F_1)<0\}
\]
is contained in the unconditional event $\{\LLR(F^\star,F_1)<0\}$.  This avoids
conditioning on transcript-dependent candidate compatibility.

Every process in $S_v(F^\star,F_1)$ is correct in the actual execution, so its
samples are genuinely distributed according to $P_{i,0}^{\otimes T}$ and are
independent across processes.  For any $s\in[0,1]$, Markov's inequality gives
\begin{align*}
    \bP_0\bigl[\LLR(F^\star,F_1)<0\bigr]
    &\le
    \bE_0\left[e^{-s\LLR(F^\star,F_1)}\right]\\
    &=
    \prod_{i\in S_v(F^\star,F_1)}
    \left(
        \sum_{x\in\mathcal{X}_i}
        p_{i,0}(x)^{1-s}p_{i,1}(x)^s
    \right)^T.
\end{align*}
Optimizing over $s$ yields
\[
    \bP_0\bigl[\LLR(F^\star,F_1)<0\bigr]
    \le
    e^{-T\cC_{S_v(F^\star,F_1)}}
    \le
    e^{-T\Lambda_v}.
\]
There are at most $N_f$ candidate sets $F_1$, so a union bound gives
\[
    \bP_0[\widehat\theta_v\neq0]
    \le
    N_f e^{-T\Lambda_v}.
\]

The hypothesis-$1$ case is symmetric.  If the true candidate is
$(1,F^\star)$, an error implies that some compatible $(0,F_0)$ wins, i.e.,
$\LLR(F_0,F^\star)\ge0$.  For any $s\in[0,1]$,
\[
    \bP_1[\LLR(F_0,F^\star)\ge0]
    \le
    \bE_1\left[e^{(1-s)\LLR(F_0,F^\star)}\right],
\]
and the right-hand side is the same Chernoff affinity product.  Optimizing over
$s$ and union-bounding over $F_0$ gives the same bound.  Taking the worst case
over actual fault sets and Byzantine strategies proves~\eqref{eq:finite-error-bound}.
\end{proof}

\begin{proposition}[Achievability by causal-state flooding]
\label{prop:achievability}
There exists a deterministic authenticated protocol family
$\boldsymbol\Pi^{\mathrm{causal}}$ that terminates in $n-1$ rounds, sends
$O(n|E|)$ correct-process transmissions, and satisfies
\[
    \cE_v(\boldsymbol\Pi^{\mathrm{causal}})\ge\Lambda_v.
\]
Consequently, $\cE_v^\star\ge\Lambda_v$.
\end{proposition}

\begin{proof}
Use the causal-state protocol and survivor decoder for every sample size $T$.
The communication phase takes $n-1$ rounds and~\eqref{eq:causal-message-count}
gives the message bound.  Lemma~\ref{lem:finite-sample-ach} gives
\[
    R_{v,T}
    \le
    \min\{1,N_f e^{-T\Lambda_v}\}.
\]
Since $n$ and $f$ are fixed as $T\to\infty$, $N_f$ is a subexponential
prefactor and therefore
\[
    \liminf_{T\to\infty}
    -\frac{1}{T}\log R_{v,T}
    \ge
    \Lambda_v.
\]
\end{proof}

\begin{proof}[Proof of Theorem~\ref{thm:main}]
Proposition~\ref{prop:converse} gives $\cE_v^\star\le\Lambda_v$, while
Proposition~\ref{prop:achievability} gives $\cE_v^\star\ge\Lambda_v$ together
with the stated round, transmission-count, and finite-sample bounds.  Hence
\[
    \cE_v^\star=\Lambda_v.
\]
\end{proof}

\subsection{Immediate consequences and scope}
\label{subsec:immediate-consequences}

Theorem~\ref{thm:main} separates the statistical role of observations from the
two-world communication obstruction.  It also makes several boundaries of the
result explicit.

\begin{corollary}[Positive-exponent criterion]
\label{cor:positive-exponent}
The target can achieve a strictly positive error exponent if and only if
\[
    \cC_{S_v(F_0,F_1)}>0
\]
for every two admissible fault sets $F_0,F_1$.  Equivalently, every candidate
pair must leave in the target pair component at least one collection of
common-correct sources that statistically distinguishes hypotheses $0$ and $1$.
\end{corollary}

\begin{remark}[Target observations]
Because the target is correct in both candidate worlds, $v\in S_v(F_0,F_1)$ for
every pair.  Thus if $P_{v,0}\neq P_{v,1}$, the target's own evidence can never
be neutralized by the $2f$-style two-world deletion that applies to other
processes.
\end{remark}

\begin{remark}[Static faults, synchrony, and authentication]
The proof uses static fault sets, synchronous deadlines, and transferable
unforgeable signatures in essential ways.  Mobile Byzantine faults, asynchronous
networks, and arbitrary correlated observations require different
characterizations and are outside the present theorem.
\end{remark}

\begin{remark}[Finite hypothesis families]
A finite multi-hypothesis extension follows by applying the same pair-graph
argument to every ordered pair of distinct hypotheses and taking the minimum
pairwise Chernoff information.  We defer the formal statement to an appendix so
that the binary case remains the conceptual core.
\end{remark}


\section{Sparse Preservation of Authenticated Pair Information}\label{app:sparse}

The characterization of Theorem~\ref{thm:main} depends on the connected
components of all authenticated pair graphs.  This section shows that these
components can be preserved on a sparse classical fault-tolerant spanner.  The
result is a reduction: the VFT-spanner constructions themselves are imported
from prior work~\cite{BodwinEtAl2018}.

\begin{definition}[Vertex-fault-tolerant spanner]
A subgraph $J\subseteq G$ is a $q$-vertex-fault-tolerant ($q$-VFT)
$t$-spanner if for every $X\subseteq V$ with $|X|\le q$ and all
$a,b\notin X$,
\[
    \dist_{J-X}(a,b)
    \le
    t\,\dist_{G-X}(a,b),
\]
with the usual convention that the inequality is vacuous when the right-hand
side is infinite.
\end{definition}

For a graph $Q$ and fault sets $F_0,F_1$, write
$H_Q(F_0,F_1)$ for the authenticated pair graph obtained from $Q$.

\begin{theorem}[VFT spanners preserve authenticated pair graphs]
\label{thm:pair-spanner}
Let $J\subseteq G$ be a $2f$-VFT $t$-spanner.  For all fault sets
$F_0,F_1$ with $|F_0|,|F_1|\le f$ and all vertices
$a,b\notin F_0\cup F_1$,
\begin{equation}
\label{eq:pair-spanner-stretch}
    \dist_{H_J(F_0,F_1)}(a,b)
    \le
    t\,\dist_{H_G(F_0,F_1)}(a,b).
\end{equation}
Consequently, $H_J(F_0,F_1)$ and $H_G(F_0,F_1)$ have identical connectivity
among vertices outside $F_0\cup F_1$.
\end{theorem}

\begin{proof}
Fix $F_0,F_1$ and let
$A_0=F_0\setminus F_1$, $A_1=F_1\setminus F_0$.
It suffices to replace every edge $\{x,y\}$ of a shortest $a$--$b$ path in
$H_G(F_0,F_1)$ by a path of length at most $t$ in $H_J(F_0,F_1)$.

For such an edge define
\[
    X=(F_0\cup F_1)\setminus\{x,y\}.
\]
Then $|X|\le 2f$, and the original edge $\{x,y\}$ survives in $G-X$.
Because $J$ is a $2f$-VFT $t$-spanner, there is an $x$--$y$ path $Q$ in
$J-X$ of length at most $t$.  Every internal vertex of $Q$ lies outside
$F_0\cup F_1$.  The endpoints may belong to $A_0$ or $A_1$, but because
$\{x,y\}$ itself belongs to the pair graph, they are not an opposite
$A_0$--$A_1$ pair.  Hence no edge of $Q$ joins an $A_0$ vertex to an $A_1$
vertex: every internal endpoint is common-correct, and the only edge whose two
endpoints could both lie in $F_0\cup F_1$ would be a direct $x$--$y$ edge.
Also, no vertex of $Q$ lies in $F_0\cap F_1$.  Therefore $Q$ is entirely
contained in $H_J(F_0,F_1)$.

Replacing the edges of the shortest pair-graph path and concatenating the
replacement paths proves~\eqref{eq:pair-spanner-stretch}.  Since $J\subseteq G$,
pair-graph connectivity in $J$ implies pair-graph connectivity in $G$; the
reverse implication follows from the finite stretch bound.
\end{proof}

\begin{corollary}[Sparse substrate preserves the exact exponent]
\label{cor:sparse-exponent}
If $J$ is a $2f$-VFT spanner of $G$, then for every correct target $v$,
\[
    \Lambda_v(J)=\Lambda_v(G),
    \qquad
    \mathcal E_v^*(J)=\mathcal E_v^*(G).
\]
For $f\ge1$, known VFT-spanner constructions give, for every fixed integer
$k\ge1$, a $(2k-1)$-stretch substrate with
\[
    O_k\!\left(f^{1-1/k}n^{1+1/k}\right)
\]
edges, up to the constant incurred by replacing $f$ by $2f$.  For $f=0$, an
ordinary $(2k-1)$-spanner suffices and has $O_k(n^{1+1/k})$ edges.  Equivalently,
the two cases can be summarized by the uniform bound
$O_k((f+1)^{1-1/k}n^{1+1/k})$.
\end{corollary}

\begin{proof}
Theorem~\ref{thm:pair-spanner} preserves, for every $(F_0,F_1)$, the target
pair component among common-correct vertices.  Hence it preserves every source
set $S_v(F_0,F_1)$ and therefore the minimum defining $\Lambda_v$.
Theorem~\ref{thm:main} then gives equality of optimal exponents.  For
$f\ge1$, the stated sparsity bound is the classical VFT-spanner upper bound of
Bodwin, Dinitz, Parter, and Vassilevska Williams~\cite{BodwinEtAl2018} applied
with fault budget $2f$.  For $f=0$, no fault tolerance is required and an
ordinary fixed-stretch spanner gives the stated $O_k(n^{1+1/k})$ bound.
\end{proof}

\begin{corollary}[Causal-state flooding on a sparse substrate]
\label{cor:sparse-messages}
Run the causal-state protocol of Theorem~\ref{thm:main} on a $2f$-VFT spanner
$J$ of $G$.  It attains the same exponent $\Lambda_v(G)$ using
\[
    O(n|E(J)|)
\]
correct-process transmissions.  With the classical fixed-stretch construction quoted above, this is
\[
    O_k\!\left(f^{1-1/k}n^{2+1/k}\right)
\]
correct-process transmissions for stretch $2k-1$ when $f\ge1$.  For $f=0$,
an ordinary spanner gives $O_k(n^{2+1/k})$ scheduled transmissions.  A uniform
statement is $O_k((f+1)^{1-1/k}n^{2+1/k})$.
\end{corollary}

\begin{remark}[Transmission metric]
Corollary~\ref{cor:sparse-messages} concerns scheduled transmissions in the
unbounded-payload model.  Appendix~\ref{app:limits} discusses bit complexity
and computation separately.
\end{remark}


\section{Assumptions and Scope}\label{app:limits}

Theorem~\ref{thm:main} is exact under the stated model.  This appendix
collects the assumptions on which the characterization depends and identifies
the dimensions that require different models or techniques.

\paragraph{Message count versus bit complexity.}
The causal-state protocol uses at most $2(n-1)|E|$ transmissions, but a state
recursively contains authenticated causal history.  Its naive serialized size
can therefore be large.  The theorem uses the unbounded-payload transmission model.  It records that
exact achievability can be witnessed by one standard causal flood rather than
by path-indexed protocol state.  A polynomial total-bit bound would require a
succinct transferable provenance mechanism or a different information-theoretic
representation, and remains a natural strengthening of the present result.

\paragraph{Information versus computation.}
Candidate compatibility and the survivor decoder are information-theoretic and
may enumerate fault sets.  Appendix~\ref{app:complexity} proves that exact computation of $\Lambda_v$ is
W[1]-hard in $f$ even for a fixed homogeneous binary source family.  This does
not prove that the full survivor decoder is W[1]-hard, nor that every exact
protocol requires hard local computation.  The main theorem is therefore an exact information boundary together with
sufficiency of a standard causal flood in the stated transmission-count model;
efficient decoding remains a separate algorithmic question.

\paragraph{Static faults and synchrony.}
The two candidate sets $F_0,F_1$ are fixed, and synchronous deadlines make a
missing correct-to-correct message observable at the receiver.  Both facts are
used in causal-path exposure.  Mobile Byzantine faults or asynchronous
communication can break the edge-by-edge certification argument and may require
a different pair object.

\paragraph{Transferable ideal authentication.}
The structural theorem uses ideal unforgeability, session/round binding, and
transferability of the concrete signed objects actually produced.  It does not
require deterministic signature bytes: signature-generation coins are
hypothesis-independent randomness and are absorbed into the common auxiliary
variable in the converse.  A computational instantiation must choose security
parameters so that authentication failure is negligible on the
$e^{-\Theta(T)}$ statistical scale if one wants to preserve the exact exponent.

\paragraph{Conditional independence.}
The converse uses product observations across processes.  For a process that is
correct in only one candidate world, the construction supplies an auxiliary
observation from the opposite hypothesis.  Conditional independence makes these
auxiliary variables independent of the common-correct source experiment and
gives them the same law in both channel representations.  With arbitrary
cross-process correlation, a different coupling or a joint information quantity
is needed.

\paragraph{Asymptotic regime.}
The network, $n$, $f$, and any finite hypothesis family are fixed while
$T\to\infty$.  Candidate-count factors are therefore subexponential in $T$.
Joint limits such as $n=n(T)$ require retaining these terms and may yield a
different exponent.

\paragraph{Temporal specialization.}
The cut theorem~\ref{thm:temporal-cut} is combinatorial and does not assume a
homogeneous sensing channel.  Homogeneity enters only when translating changed
reporter count into the closed-form statistical exponent
$c[d_{\rm rel}^{\pi}(\Gamma)-2f]_+$.  For heterogeneous or
magnitude-sensitive trace channels, Theorem~\ref{thm:main} still applies to each
pair of global orders, but the remaining optimization becomes weighted by the
full Chernoff affinity profiles rather than by reporter count alone.

\section{Computational Complexity of Exact Pair Information}
\label{app:complexity}

The main theorem is an exact information characterization, but evaluating its
minimum can itself be computationally hard.  We show W[1]-hardness parameterized
by the Byzantine budget $f$ even for a restricted source family in which every
informative process carries the same fixed binary experiment.

For an informative source set $D\subseteq V\setminus\{v\}$, define the maximum
number of informative processes suppressed by a pair of fault explanations:
\begin{equation}
\label{eq:phi-def}
\Phi_v(D,f)
:=
\max_{\substack{F_0,F_1\subseteq V\setminus\{v\}\\|F_0|,|F_1|\le f}}
\left|D\setminus S_v(F_0,F_1)\right|.
\end{equation}

Fix once and for all a rational $p\in(0,1/2)$.  Give every $i\in D$ the same
binary experiment
$P_{i,0}=\operatorname{Bern}(p)$ and
$P_{i,1}=\operatorname{Bern}(1-p)$, and give every process outside $D$
hypothesis-independent observations.  The informative sources share the same
Chernoff profile, with fixed constant
\begin{equation}
\label{eq:fixed-c-hardness}
c=-\log\!\left(2\sqrt{p(1-p)}\right)>0.
\end{equation}
Hence for every source set $S$,
$\mathcal C_S=c|S\cap D|$, and therefore
\begin{equation}
\label{eq:lambda-phi-relation}
\frac{\Lambda_v}{c}
=|D|-\Phi_v(D,f).
\end{equation}
The normalized value $\Lambda_v/c$ is thus an integer on this fixed source
family, avoiding any issue about representing the logarithmic constant $c$.

\begin{theorem}[Exact pair information is \Wone-hard]
\label{thm:w1-hard}
Computing the normalized value $\Lambda_v/c$, and hence computing
$\Lambda_v$ exactly, is W[1]-hard parameterized by $f$, even for the fixed
binary source family above and bipartite communication graphs in which every
informative process has degree two.
\end{theorem}

\begin{proof}
We reduce from $k$-\textsc{Clique}, which is W[1]-complete
\cite{DowneyFellows1999}.  Let $Q=(U,E_Q)$ and $k$ be a clique instance and set
$M:=2k+1$.  Construct a bipartite communication graph with target $v$ and, for
every $u\in U$, a selector process $x_u$ adjacent to $v$.  For every edge
$e=\{u,w\}\in E_Q$, create $M$ informative degree-two processes
$d_{e,1},\ldots,d_{e,M}$, each adjacent exactly to $x_u$ and $x_w$.  Let $D$ be
the set of all these informative processes and set $f=k$.

For arbitrary candidate sets $F_0,F_1$, let
$B=\{u\in U:x_u\in F_0\cap F_1\}$.  Consider an informative process
$d_{e,j}\notin F_0\cup F_1$ for $e=\{u,w\}$.  If
$x_u\notin F_0\cap F_1$, then the path $d_{e,j}-x_u-v$ survives the pair graph:
$d_{e,j}$ and $v$ are common-correct, and an edge from a common-correct process
to a selector faulty in only one world is not an opposite-exclusive edge.
Thus $d_{e,j}$ is pair-disconnected from $v$ exactly when both endpoint
selectors $x_u,x_w$ are common faults.  Non-direct suppression therefore
contributes $M|E_Q[B]|$, while at most
$|F_0\cup F_1|\le2k=M-1$ further informative processes can be suppressed by
corrupting them directly.

If $Q$ contains a $k$-clique $K$, choose
$F_0=F_1=\{x_u:u\in K\}$; this suppresses $M\binom{k}{2}$ informative
processes.  If $Q$ has no $k$-clique, every $B$ of size at most $k$ spans at
most $\binom{k}{2}-1$ edges, and every pair attack suppresses at most
\[
M\left(\binom{k}{2}-1\right)+(M-1)
=M\binom{k}{2}-1.
\]
Let $\kappa:=M\binom{k}{2}$.  We have
\[
Q\text{ has a }k\text{-clique}
\quad\Longleftrightarrow\quad
\Phi_v(D,k)\ge \kappa
\quad\Longleftrightarrow\quad
\frac{\Lambda_v}{c}\le |D|-\kappa.
\]
The final equivalence is~\eqref{eq:lambda-phi-relation}; the normalized
threshold is an integer computable from the clique instance.  The construction
maps the parameter to $f=k$, proving W[1]-hardness of $\Lambda_v/c$ and
therefore of exact $\Lambda_v$ evaluation.
\end{proof}

\begin{remark}[An exact XP algorithm]
The definition of $\Lambda_v$ can be evaluated by enumerating all admissible
pairs $(F_0,F_1)$, constructing the pair graph, finding $S_v(F_0,F_1)$, and
evaluating its product Chernoff information.  With
$N_f=\sum_{j=0}^{f}\binom{n-1}{j}$ candidate fault sets, the graph part takes
$O(N_f^2(n+|E|))=n^{2f+O(1)}$ time, plus the cost of the source-information
evaluations.  Thus exact evaluation is in XP for fixed $f$ in the finite-source
models considered here, while Theorem~\ref{thm:w1-hard} rules out an exact FPT
algorithm for the restricted family above under FPT $\ne$ W[1].  This does not
imply that every decoder or every exact protocol requires hard local
computation.
\end{remark}

\section{Finite-Hypothesis Identification and Composite Testing}
\label{app:multi}

The binary theorem has two distinct finite-hypothesis extensions:
\emph{identifying} one state from a finite family requires distinguishing every
pair of states, whereas testing one distinguished baseline against a composite
alternative does not require the alternatives to be distinguishable from one
another.

\begin{proposition}[Finite-hypothesis identification]
\label{prop:finite-identification}
Let the global state range over a fixed finite family $\Theta$, and suppose the
target must output the exact state $\theta\in\Theta$.  For distinct
$\theta,\theta'$, let $\mathcal C_S(\theta,\theta')$ be the Chernoff information
between the corresponding product observations of $S$.  Then the optimal
minimax identification exponent is
\begin{equation}
\label{eq:finite-identification}
\mathcal E_{v,\mathrm{id}}^*
=
\min_{\substack{\theta\ne\theta'\\|F|,|F'|\le f}}
\mathcal C_{S_v(F,F')}(\theta,\theta').
\end{equation}
\end{proposition}

\begin{proof}
For the converse, restrict the identification problem to any ordered pair
$\theta\ne\theta'$ and any two admissible fault sets; the two-world simulation
of Proposition~\ref{prop:converse} gives the corresponding pairwise upper
bound.  For achievability, run the same causal-state protocol and call a compatible
candidate $(\theta,F)$ a survivor if it wins every direct comparison against
compatible candidates with a different state, using the observations exposed
from $S_v(F,F')$.  Survivors with different states cannot coexist because their
direct comparison has a unique winner; output the survivor state if one exists.
If the true candidate wins all comparisons it is a survivor, so an error implies
one pairwise loss by the true candidate.  A union bound over the fixed finite
sets of states and fault sets achieves~\eqref{eq:finite-identification}; its
prefactor is subexponential in $T$.
\end{proof}

Now fix one distinguished baseline $\theta_0$ and a nonempty finite alternative
set $\mathcal A\subseteq\Theta\setminus\{\theta_0\}$.  The target outputs only a
binary answer: baseline or alternative.  Define
$\mathcal E_{v,\theta_0:\mathcal A}^*$ using worst-case error over
$\theta=\theta_0$ and all $\theta\in\mathcal A$, together with the same fault
and Byzantine-strategy maxima as in Definition~\ref{def:risk}.

\begin{proposition}[Baseline versus a finite composite alternative]
\label{prop:finite-composite}
For the binary decision problem above,
\begin{equation}
\label{eq:finite-composite}
\mathcal E_{v,\theta_0:\mathcal A}^*
=
\min_{\substack{\theta\in\mathcal A\\|F_0|,|F_1|\le f}}
\mathcal C_{S_v(F_0,F_1)}(\theta_0,\theta).
\end{equation}
In particular, no Chernoff term comparing two members of $\mathcal A$ appears.
\end{proposition}

\begin{proof}
The converse follows by restricting the composite problem to any single
alternative $\theta\in\mathcal A$ and applying the binary two-world converse.
For achievability, compare only compatible baseline candidates
$(\theta_0,F_0)$ with compatible alternative candidates $(\theta,F_1)$,
$\theta\in\mathcal A$, and output the baseline iff some baseline candidate
wins all such comparisons.  Under the baseline, an error implies that the true
baseline candidate loses one comparison; under a true alternative, an erroneous
baseline output implies that some baseline candidate beats the true alternative
candidate.  The binary Chernoff bound and a finite union bound give
\eqref{eq:finite-composite}.  Alternative--alternative comparisons are neither
needed nor present in the exponent.
\end{proof}


\end{document}